\documentclass[aps,pre,preprint,superscriptaddress]{revtex4-1}

\usepackage[pdftex]{graphicx}
\usepackage[inkscapelatex=false]{svg}%
\usepackage{dcolumn}
\usepackage{bm}
\usepackage{mathtools}
\newtagform{Eq}{Eq.}{}
\usetagform{Eq}

\begin{document}
\preprint{The original paper was accepted for publication in Physical Review Letters on 3 October 2024. This extended}
\preprint{version combines the content of the original paper and the Supplementary Materials for the benefit of readers.}

\title{Discovery of Electromagnetic Surface Waves at the Interface Between Perfect Electric Conductor and Perfect Magnetic Conductor Parallel-Plate Waveguides}

\author{Seong-Han Kim}
\author{Chul-Sik Kee}\email{cskee@gist.ac.kr}
\affiliation{Advanced Photonics Research Institute, GIST, Gwangju 61005, Republic of Korea}

\date{\today}

\begin{abstract}
We propose new electromagnetic surface waves at the interface formed by connecting a perfect electric conductor (PEC) and a perfect magnetic conductor (PMC) parallel plate waveguides containing materials with positive permittivities and permeabilities. This challenges the conventional understanding that surface waves require materials with negative permittivity or permeability. Theoretical mode analysis and numerical simulations have confirmed the existence of surface waves at the PEC-PMC interface. Additionally, a simulated prism coupling experiment validated the excitation of the surface wave at the PEC-PMC interface. The resonant response of the localized surface waves on the enclosed PEC-PMC surface of a cylinder also closely resembles that of a Drude cylinder. Our finding broadens the understanding of the conditions for generating electromagnetic surface waves and deepens our comprehension of electromagnetic phenomena.
\end{abstract}


\maketitle

\section{Introduction}
Electromagnetic (EM) surface waves occur at the interface between two materials when the product of their permittivities or permeabilities is negative.\cite{sw1} A typical example is the surface plasmon, which represents the surface wave occurring at the interface between a metal and a dielectric. Surface plasmons are crucial in manipulating and confining light at the nanoscale, facilitating various applications like nanoantennas, surface-enhanced spectroscopy, and subwavelength imaging.\cite{sw3,sw4} Furthermore, surface plasmon resonance sensors utilize surface waves to detect changes in the refractive index near the sensor surface, enabling applications in biosensing, environmental monitoring, and chemical analysis.\cite{sw2} Surface waves also play a significant role in improving the performance and design of antennas.\cite{sw5} A deeper comprehension of surface wave behavior is key to enhancing antenna designs, resulting in better efficiency and increased bandwidth. 

A perfect electric conductor (PEC) surface fully reflects an incident wave with a 180-degree reflection phase due to a zero electric field on its surface, while a perfect magnetic conductor (PMC) surface introduces a zero-degree phase shift because the magnetic field is zero on its surface. Only a transverse electromagnetic (TEM) mode can propagate if the distance between plates is less than half of a wavelength in a parallel plate waveguide. Connecting the PEC and the PMC waveguides forms a vertical virtual electric/magnetic wall, preventing a TEM mode transfer.\cite{pmc1} When a TEM mode encounters the interface, it faces either a virtual electric wall from the PMC waveguide or a virtual magnetic wall from the PEC waveguide, resulting in reflection.

Recent numerical simulations have further demonstrated that objects enclosed by a virtual pillar magnetic wall can be effectively shielded from electromagnetic waves.\cite{pmc3} Moreover, the properties of two dimensional (2D) photonic crystals composed of the pillar magnetic walls exhibit significant similarity to those of 2D photonic crystals made of infinitely long perfect electric conductor cylinders.\cite{pmc2} This similarity arises from the fact that the scattering properties of the pillar magnetic wall for a TEM mode closely resemble those of an infinitely long metallic cylinder for a transverse magnetic (TM) mode. The observed similarity suggests that a surface wave propagating along the interface between parallel PEC and PMC plate waveguides can indeed exist, resembling a surface wave propagating along the interface between a conductor and a dielectric.

\begin{figure}[htbp]
 \centering %
\includegraphics[width=76mm]{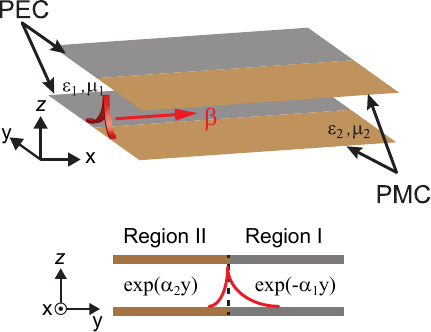}%
\caption{A schematic of a connected PEC and PMC parallel plate waveguide with a separation of the plates denoted as $h$, and a surface wave propagating along the interface of a PEC and a PMC parallel plate waveguides with a wave vector, $\beta$. Region I (Region II) denotes a PEC (PMC) parallel plate waveguide region. A material with positive permittivity $\epsilon_{1}$ ($\epsilon_{2}$) and permeability $\mu_{1}$ ($\mu_{2}$) is contained within a PEC (PMC) parallel plate waveguide. \label{fig1} }
\end{figure}

In this paper, we demonstrate the presence of surface waves at the interface between a PEC and a PMC parallel plate waveguides, which contain materials with positive permittivities and permeabilities. Theoretical mode analysis based on boundary conditions was conducted to derive the dispersion relation of the surface wave, closely matching the simulated dispersion relation obtained from the finite-element method (FEM). Additionally, a simulated prism coupling experiment verified the excitation of the surface wave at the interface of a PEC and a PMC parallel plate waveguides. We also show how the resonant response of the localized surface waves on the enclosed PEC-PMC surface of a 2D cylinder closely resembles that of a Drude cylinder.


\section{Results}

\subsection{Mathematical derivation of the dispersion relation of the surface wave at the PEC-PMC interface through a modal expansion and boundary conditions}
  
Figure 1 illustrates the schematic of a connected PEC and PMC parallel plate waveguide with a separation of the plates denoted as $h$. Region I (Region II) denotes a PEC (PMC) parallel plate waveguide region. A material with positive permittivity $\epsilon_{1}$ ($\epsilon_{2}$) and permeability $\mu_{1}$ ($\mu_{2}$) is contained within a PEC (PMC) parallel plate waveguide.  The interface between PEC and PMC parallel plate waveguides is termed a PEC-PMC interface for convenience. A surface wave at the PEC-PMC interface propagates along the $x$-axis with a wave vector, $\beta$, and the fields should decay in $\pm y$-direction. To obtain the dispersion relation of the surface waves, we applied a theoretical framework based on a modal expansion of EM fields. 
  
\subsubsection{Wave equations}
We assume time-harmonic electric and magnetic fields with an $\exp{(-i\omega t)}$ dependence and wave propagation along x-direction with propagation constant $\beta$.

\begin{subequations}
\label{eq01}
\begin{align}
    \mathcal{\vec{E}}(\mathbf{r},t)=Re{[\textbf{E}(y,z)\exp{[i(\beta x-\omega t)]}]},\\
\mathcal{\vec{H}}(\mathbf{r},t)=Re{[\textbf{H}(y,z)\exp{[i(\beta x-\omega t)]}]}.
\end{align}
\end{subequations}
For the Cartesian components of $\mathbf{E}$ and $\mathbf{H}$, Maxwell's curl equations become
\begin{subequations}
  \label{eq02}
\begin{align}
   \frac{\partial E_{z}} {\partial y}- \frac{\partial E_{y}} {\partial z} &= i\omega \mu H_{x}, \label{eq02:a} \\
    \frac{\partial E_{x}} {\partial z}- i \beta E_{z} &= i\omega \mu H_{y}, \label{eq02:b} \\
    i \beta E_{y}-  \frac{\partial E_{x}} {\partial y} &= i \omega \mu H_{z}, \label{eq02:c}\\
    \frac{\partial H_{z}} {\partial y}- \frac{\partial H_{y}} {\partial z} &= -i\omega \epsilon E_{x}, \label{eq02:d} \\
    \frac{\partial H_{x}} {\partial z}- i \beta H_{z} &= -i\omega \epsilon E_{y},  \label{eq02:e}\\
    i \beta H_{y}-  \frac{\partial H_{x}} {\partial y} &= -i \omega \epsilon E_{z}.\label{eq02:f}
\end{align}
\end{subequations}
From Eqs. \eqref{eq02:a}-\eqref{eq02:f}, transverse components are expressed in terms of $E_{x}$ and $H_{x}$ as 
\begin{subequations}
  \label{eq03}
\begin{align}
    H_{y} &=\frac{i}{\beta_{t}^2}\left[\beta \frac{\partial H_{x}} {\partial y} - \omega \epsilon \frac{\partial E_{x}} {\partial z}   \right],   \label{eq03:a} \\
    H_{z} &=\frac{i}{\beta_{t}^2}\left[\beta \frac{\partial H_{x}} {\partial z} + \omega \epsilon \frac{\partial E_{x}} {\partial y}   \right],   \label{eq03:b} \\
    E_{y} &=\frac{i}{\beta_{t}^2}\left[\beta \frac{\partial E_{x}} {\partial y} + \omega \mu \frac{\partial H_{x}} {\partial z}   \right],   \label{eq03:c}\\
    E_{z} &=\frac{i}{\beta_{t}^2}\left[\beta \frac{\partial E_{x}} {\partial z} - \omega \mu \frac{\partial H_{x}} {\partial y}   \right],   \label{eq03:d}
\end{align}
\end{subequations}
where $\beta_{t}^2=n^2 k_{0}^2-\beta^2$, $ k_{0}^2=\omega^2 \epsilon_{0} \mu_{0}$ and $n=\sqrt{\mu_r\epsilon_r}$ is the refractive index of the medium.
Substituting Eqs. \eqref{eq03:a} and \eqref{eq03:b} into \eqref{eq02:d}, we obtain 
\begin{equation}
\label{eq04}
  \left( \frac{\partial^{2}}{\partial y^{2}} +\frac{\partial^{2}}{\partial z^{2}}+ \beta_t^{2} \right)H_{x} =0.
\end{equation}
and by substituting Eqs. \eqref{eq03:c} and \eqref{eq03:d} into \eqref{eq02:a}, we have 
\begin{equation}
\label{eq05}
  \left( \frac{\partial^{2}}{\partial y^{2}} +\frac{\partial^{2}}{\partial z^{2}}+ \beta_t^{2} \right)E_{x} =0.
\end{equation}
\subsubsection{Mode solutions}
Eq.\eqref{eq04} and \eqref{eq05} can be solved by the method of separation of variables by letting $H_{x}(\text{or}\enspace E_{x})=Y(y)Z(z)$. By substituting this into Eq. \eqref{eq04} and \eqref{eq05}, we obtain 
\begin{equation}
\label{eq06}
  \frac{1}{Y}\frac{\partial^{2}Y}{\partial y^{2}} +\frac{1}{Z}\frac{\partial^{2}Z}{\partial z^{2}}+ \beta_t^{2} =0.
\end{equation}
Here, each term in Eq.\eqref{eq06} should be a constant, so we define separation constants $k_{y}$ and $k_{z}$, such that
  \begin{align}
  \frac{\partial^{2}Y}{\partial y^{2}}+ k_y^{2}Y &=0,\label{eq07} \\
   \frac{\partial^{2}Z}{\partial z^{2}}+ k_z^{2}Z &=0,\label{eq08}
\end{align}
where 
\begin{equation}
\label{eq09}
  k_{y}^2+k_{z}^2 =\beta_{t}^2.
\end{equation}

The solutions of the above two equations can be written as
\begin{align}
    &Y(y) = \begin{cases}  A e^{-|k_{y}|y} + Be^{+|k_{y}|y}, \quad \text{for imaginary $k_{y}$}, \\
    A'\cos(k_{y}y)+B'\sin(k_{y}y),\quad \text{for real $k_{y}$},
    \end{cases}\label{eq10}\\
    &Z(z)=\begin{cases}
    \cos(k_{z}z) \\
    \sin(k_{z}z),\label{11}
    \end{cases}
    \end{align}
where $A$,$B$, $A'$, and $B'$ are arbitrary constant.\\
Consider an interface between PEC and PMC parallel plate waveguides as depicted in Fig. \ref{fig1}. PEC and PMC regions are denoted by subscripts "I" and "II", respectively. Firstly, solutions are obtained independently for both regions and then connected later at the interface between the two regions to satisfy boundary conditions. 
For an EM wave to propagate along the interface, the fields should decay in the $\pm y$-directions (no propagation mode in the $y$-direction). Thus the field function $Y(y)$ should take the form of the upper line in equation \eqref{eq10}. In the PEC region ($y>0$), the second term should be discarded ($B=0$), and in the PMC region, the first term should be discarded ($A=0$). This means that $k_y$ must be imaginary in both regions (\textit{i.e.} $k_{y} =\pm i\sqrt{\beta^2-n^{2}k_0^{2}+k_{z}^2}$ ). Here, we can define decay constants as 
\begin{equation}\label{eq12}
    \alpha_{j} \equiv \sqrt{\beta^2-n_{j}^{2}k_0^{2}+k_{j,z}^2}, \quad j=1,2.
\end{equation} 
Also, the boundary conditions that should be satisfied in each region are 
\begin{align}
    &\textbf{n}\times\textbf{E}=0 \quad \text{at PEC surfaces},\label{eq13}\\
    &\textbf{n}\times\textbf{H}=0 \quad \text{at PMC surfaces},\label{eq14}
    \end{align}
where $\textbf{n}$ is a normal vector to surfaces.\
Under the above conditions, we assume that there are two sets of solutions: TE and TM modes. The TE modes are characterized by fields with a nonzero magnetic field ($H_x \neq 0$) and a zero electric field ($E_x = 0$) in the propagation direction. Conversely, the TM modes have a nonzero electric field ($E_x \neq 0$) and a zero magnetic field ($H_x = 0$) in the propagation direction.
After applying the boundary conditions \eqref{eq13} and \eqref{eq14}, TE solutions are given in which 
\begin{align}
    &E_x = \begin{cases}  A e^{-|k_{1,y}|y}\sin(k_{1,z}z) \quad \text{for $y\geq0$ }\\
    B e^{|k_{2,y}|y}\cos(k_{2,z}z) \quad \text{for $y<0$ },
    \end{cases}\label{eq15}\\
    & H_x=0, \label{eq16}
    \end{align}
where 
\begin{equation*}
    |k_{1,y}|\equiv\alpha_{1,m}=\sqrt{\beta^2-n_1^{2} k_{0}^2+(m\pi/h)^2},
\end{equation*}
, $k_{1,z}=m\pi/h$, $m=1,2,3,\ldots$, and
$k_{2,y}=\alpha_{2,n}$, $k_{2,z}=n\pi/h$, $n=0,1,2,3,\ldots$.
Similarly, the TM solutions can be written as 
\begin{align}
    &H_x = \begin{cases}  C e^{-|k_{1,y}|y}\cos(k_{1,z}z) \quad \text{for $y\geq0$ }\\
    D e^{|k_{2,y}|y}\sin(k_{2,z}z) \quad \text{for $y<0$ },
    \end{cases}\label{eq17}\\
    & E_x=0, \label{eq18}
    \end{align}
where $k_{1,y}=\alpha_{1,p}$, $k_{1,z}=p\pi/h$, $p=0,1,2,3,\ldots$, and $k_{2,y}=\alpha_{2,q}$, $k_{2,z}=q\pi/h$, $q=1,2,3,\ldots$.\\
In these solutions, $A$, $B$, $C$, and $D$ are arbitrary amplitude constants.
It's important to note that while TE and TM modes alone do not exist in this system, but rather a surface mode exists as a combination of TE and TM modes, known as a hybrid mode. Nevertheless, TE and TM modes are useful for deriving the hybrid mode. \\
We obtain the general expression of the field components, including all possible modes (the TE, TM, and hybrid modes), derived by merging Eq. \eqref{eq15}-\eqref{eq18} and using Eq. \eqref{eq03:a}-\eqref{eq03:d}. \\
In region I,\\
\begin{align}
 E_{x}^{I} =& \sum\limits_{m=1}^{\infty}A_{k} e^{-\alpha_{1,m}y}\sin(\frac{m\pi}{h}z),\label{eq19}\\
 E_{y}^{I} =&\sum\limits_{m=1}^{\infty} \frac{-i\beta }{\beta_{1,t}^2}\alpha_{1,m}A_{m}e^{-\alpha_{1,m}y}\sin(\frac{m\pi}{h}z)-\sum\limits_{p=0}^{\infty}\frac{i\omega \mu_{1}}{\beta_{1,t}^2} \frac{p\pi}{h}C_{p}e^{-\alpha_{1,p}y}\sin(\frac{p\pi}{h}z), \label{eq20}\\
E_{z}^{I} =& \sum\limits_{m=1}^{\infty} \frac{i\beta}{\beta_{1,t}^2} \frac{m\pi}{h} A_{m}e^{-\alpha_{1,m}y}\cos(\frac{m\pi}{h}z)
+\sum\limits_{p=0}^{\infty}\frac{i\omega \mu_{1}}{\beta_{1,t}^2} \alpha_{1,p} C_{p}e^{-\alpha_{1,p}y}\cos(\frac{p\pi}{h}z),\label{eq21}\\
H_{x}^{I} =& \sum\limits_{p=0}^{\infty}C_{p} e^{-\alpha_{1,p}y}\cos(\frac{p\pi}{h}z),\label{eq22}\\
H_{y}^{I} =&\sum\limits_{m=1}^{\infty} \frac{-i\omega \epsilon_{1}}{\beta_{1,t}^2} \frac{m\pi}{h} A_{m}e^{-\alpha_{1,m}y}\cos(\frac{m\pi}{h}z)-\sum\limits_{p=0}^{\infty}\frac{i\beta}{\beta_{1,t}^2}\alpha_{1,p} C_{p}e^{-\alpha_{1,p}y}\cos(\frac{p\pi}{h}z), \label{eq23}\\
H_{z}^{I} =&\sum\limits_{m=1}^{\infty} \frac{-i\omega \epsilon_{1}}{\beta_{1,t}^2}\alpha_{1,m} A_{m}e^{-\alpha_{1,m}y}\sin(\frac{m\pi}{h}z)-\sum\limits_{p=1}^{\infty}\frac{i\beta}{\beta_{1,t}^2}\frac{p\pi}{h} C_{p}e^{-\alpha_{1,p}y}\sin(\frac{p\pi}{h}z)\label{eq24}.
\end{align}
In region II,
\begin{align}
E_{x}^{II}=&\sum\limits_{n=0}^{\infty}B_{n} e^{\alpha_{2,n}y}\cos(\frac{n\pi}{h}z),\label{eq25}\\
E_{y}^{II} =&\sum\limits_{n=0}^{\infty} \frac{i\beta}{\beta_{2,t}^2}\alpha_{2,n} B_{n}e^{\alpha_{2,n}y}\cos(\frac{n\pi}{h}z)+ \sum\limits_{q=1}^{\infty}\frac{i\omega \mu_{2}}{\beta_{2,t}^2} \frac{q\pi}{h}D_{q}e^{\alpha_{2,q}y}\cos(\frac{q\pi}{h}z), \label{eq26}\\
E_{z}^{II} =& \sum\limits_{n=0}^{\infty}\frac{-i\beta}{\beta_{2,t}^2}\frac{n\pi}{h} B_{n}e^{\alpha_{2,n}y}\sin(\frac{n\pi}{h}z)-\sum\limits_{q=1}^{\infty}\frac{i\omega \mu_{2}}{\beta_{2,t}^2} \alpha_{2,q} D_{q}e^{\alpha_{2,q}y}\sin(\frac{q\pi}{h}z),\label{eq27}\\
 H_{x}^{II}=&\sum\limits_{q=1}D_{n} e^{\alpha_{2,q}y}\sin(\frac{q\pi}{h}z),\label{eq28}\\
H_{y}^{II} =& \sum\limits_{n=0}^{\infty}\frac{i\omega \epsilon_{2}}{\beta_{2,t}^2} \frac{n\pi}{h}B_{n}e^{\alpha_{2,n}y}\sin(\frac{n\pi}{h}z)+\sum\limits_{q=1}^{\infty}\frac{i\beta}{\beta_{2,t}^2}\alpha_{2,q} D_{q}e^{\alpha_{2,q}y}\sin(\frac{q\pi}{h}z), \label{eq29}\\
H_{z}^{II} =& \sum\limits_{n=0}^{\infty}\frac{i\omega \epsilon_{2}}{\beta_{2,t}^2} \alpha_{2,n} B_{n}e^{\alpha_{2,n}y}\cos(\frac{n\pi}{h}z)+\sum\limits_{q=1}^{\infty}\frac{i\beta}{\beta_{2,t}^2}\frac{q\pi}{h} D_{q}e^{\alpha_{2,q}y}\cos(\frac{q\pi}{h}z).\label{eq30}
\end{align} 
\subsubsection{Dispersion relations of surface modes}
The dispersion relations of the surface modes are determined by the boundary conditions at the PEC-PMC interface $y=0$, 
\begin{subequations}\label{eq31}
\begin{align}
    E_{x}^{I} &= E_{x}^{II} , \label{eq31:a}\\
    E_{z}^{I} &= E_{z}^{II} ,\label{eq31:b}\\
    \epsilon_{1} E_{y}^{I} &= \epsilon_{2}E_{y}^{II} ,\label{eq31:c}\\ 
    H_{x}^{I}&= H_{x}^{II} ,\label{eq31:d}\\
    H_{z}^{I} &= H_{z}^{II} , \label{eq31:e}\\
    \mu_{1}H_{y}^{I} &= \mu_{2}H_{y}^{II}.\label{eq31:f} 
\end{align}
\end{subequations}
Since modes in one region can couple to modes in the other region, we need to consider all modes in each region when applying boundary conditions. Substituting Eq.~\eqref{eq19}-\eqref{eq30} into the boundary conditions (eq. \eqref{eq31}), we obtain the following equations:
\begin{align}
&\sum\limits_{m=1}^{\infty} A_{m} \sin(\frac{m\pi}{h}z)
  =\sum\limits_{n=0}^{\infty}B_{n}\cos(\frac{n\pi}{h}z),\label{eq32}\\ 
&\sum\limits_{p=0}^{\infty} C_{p} \cos(\frac{p\pi}{h}z)
        =\sum\limits_{q=1}^{\infty}D_{q}\sin(\frac{q\pi}{h}z),\label{eq33}\\
&\sum\limits_{m=1}^{\infty} \frac{i\beta m\pi}{\beta_{1,t}^2 h} A_{m}\cos(\frac{m\pi}{h}z)+\sum\limits_{p=0}^{\infty} \frac{i\omega \mu_{1}\alpha_{1,p}}{\beta_{1,t}^2} C_{p}\cos(\frac{p\pi}{h}z) \nonumber \\
&=\sum\limits_{n=1}^{\infty} \frac{-i\beta n\pi}{\beta_{2,t}^2 h} B_{n}\sin(\frac{n\pi}{h}z)-\sum\limits_{q=1}^{\infty} \frac{i\omega \mu_{2}\alpha_{2,q}}{\beta_{2,t}^2} D_{q}\sin(\frac{q\pi}{h}z),\label{eq34}\\
&\sum\limits_{m=1}^{\infty}\frac{-i\omega \epsilon_{1}\alpha_{1,m}}{\beta_{1,t}^2} A_{m}\sin(\frac{m\pi}{h}z)-\sum\limits_{p=1}^{\infty}\frac{i\beta p\pi}{\beta_{1,t}^2 h} C_{p}\sin(\frac{p\pi}{h}z)\nonumber \\
&=\sum\limits_{n=0}^{\infty}\frac{i\omega \epsilon_{2}\alpha_{2,n}}{\beta_{2,t}^2} B_{n}\cos(\frac{n\pi}{h}z)+\sum\limits_{q=1}^{\infty}\frac{i\beta q\pi}{\beta_{2,t}^2 h} D_{q}\cos(\frac{q\pi}{h}z).\label{eq35}
\end{align}
Equations \eqref{eq32}-\eqref{eq35} constitute four infinite sets of linear equations for the modal coefficients $A_m$, $B_n$, $C_p$, and $D_q$. Multiplying equations \eqref{eq32} and \eqref{eq33} by $\cos(l\pi z/h)$, integrating from $z=0$ to $h$, and using orthogonality relations yields
\begin{align}
 \sum\limits_{m=1}^{\infty}A_{m}\frac{2h}{\pi}I_{ml}
 &=\sum\limits_{n=0}^{\infty}B_{n}\frac{h}{2}\delta_{nl},  \enspace \text{for odd}\enspace n+l, \label{eq36}\\
 \sum\limits_{p=0}^{\infty}C_{p}\frac{h}{2}\delta_{pl}
 &=\sum\limits_{q=1}^{\infty}D_{q}\frac{2h}{\pi}I_{ql},  \quad \text{for odd}\quad q+l,\label{eq37}
 \end{align}
 where $\delta_{lp}$ is the Kronecker delta function and
\begin{align}\label{eq40}
    I_{ml}&\equiv\frac{\pi}{2h} \int_{0}^{h}\sin(m\pi z/h) \cos(l\pi z/h) \mathrm{d} z = \frac{m}{m^2-l^2}, \quad \text{for odd $m+l$}. 
\end{align}
By multiplying Eqs. \eqref{eq34} and \eqref{eq35} by $\sin(l\pi z/h)$ and integrating from $z=0$ to $h$ with orthogonality conditions, we obtain 
\begin{align}
 &\sum\limits_{m=1}^{\infty}A_{m}\frac{i\beta}{\beta_{1,t}^2}\frac{m\pi}{h}\frac{2h}{\pi}\alpha_{1,m}I_{lm}+\sum\limits_{p=0}^{\infty}C_{p}\frac{i\omega \mu_{1}}{\beta_{1,t}^2}\frac{2h}{\pi}I_{lp} 
 =\sum\limits_{n=0}^{\infty}B_{n}\frac{-i\beta}{\beta_{2,t}^2}\frac{h}{2}\frac{n\pi}{h}\delta_{nl}+\sum\limits_{q=1}^{\infty}D_{q}\frac{-i\omega \mu_{2}}{\beta_{2,t}^2}\frac{h}{2}\alpha_{2,q}\delta_{ql}, \nonumber \\
&\quad \text{for odd $m+l$ and odd $q+l$},\label{eq38} \\
 &\sum\limits_{m=1}^{\infty}A_{m}\frac{-i\omega \epsilon_{1}}{\beta_{1,t}^2}\alpha_{1,m}\frac{h}{2}\delta_{ml}+\sum\limits_{p=0}^{\infty}C_{p}\frac{-i\beta}{\beta_{1,t}^2}\frac{p\pi}{h}\frac{h}{2}\delta_{pl}
 =\sum\limits_{n=0}^{\infty}B_{n}\frac{i\omega \epsilon_{2}}{\beta_{2,t}^2}\alpha_{2,n}\frac{n\pi}{h}I_{ln}+\sum\limits_{q=1}^{\infty}D_{q}\frac{i\beta}{\beta_{2,t}^2}\frac{q\pi}{h}\frac{2h}{\pi}I_{lq}, \nonumber \\
 &\quad \text{for odd $n+l$ and odd $q+l$}.\label{eq39}
 \end{align}

It is noted that the above equations can be split into two independent sets. One set comprises even orders of $A_{m}$ and $C_{p}$, and odd orders of $B_{n}$ and $D_{q}$, while the other set comprises odd orders of $A_{m}$ and $C_{p}$, and even orders of $B_{n}$ and $D_{q}$.
For numerical calculations, considering $N$ terms in the equations produces $N$ linear equations for the first $N$ coefficients, $A_{m}$, $B_{n}$, $C_{p}$, and $D_{q}$. Numerical accuracy can be improved by using larger values of $N$.
The former set of equations is written as
\begin{align}
   \sum\limits_{\text{even}\; m}^{2N}A_{m}I_{ml}\frac{4}{\pi}&= B_{l},\enspace l = 1,3,\ldots,2N-1,\label{eq41} \\
    \sum\limits_{\text{odd}\; q}^{2N-1}D_{q}I_{ql}\frac{4}{\pi}&=C_{0}\delta_{l0}+C_{l},\enspace l = 0,2, \ldots,2N,\label{eq42}\\
   \sum\limits_{\text{even}\; m}^{2N}A_{m}I_{lm}\frac{\beta}{\beta_{1,t}^2}\frac{m\pi}{h}\frac{4}{\pi}+
    \sum\limits_{\text{even}\; p=0}^{2N}C_{p}I_{lp}\frac{\omega \mu_{1} }{\beta_{1,t}^2}\alpha_{1,p}\frac{4}{\pi}
   &=-B_{l}\frac{\beta}{\beta_{2,t}^2}\frac{l\pi}{h}-D_{l}\frac{\omega \mu_{2} }{\beta_{2,t}^2}\alpha_{2,l},\enspace l=1,3, \ldots,2N-1\label{eq43}\\
  \sum\limits_{\text{odd}\; n}^{2N-1}B_{n}I_{ln}\frac{\omega\epsilon_{2}}{\beta_{2,t}^2}\alpha_{2,n}\frac{4}{\pi}+\sum\limits_{\text{odd}\; q}^{2N-1}D_{q}I_{lq}\frac{\beta}{\beta_{2,t}^2}\frac{q\pi}{h}\frac{4}{\pi}
  &=-A_{l}\frac{\omega \epsilon_{1}}{\beta_{1,t}^2}\alpha_{1,l}-C_{l}\frac{\beta}{\beta_{1,t}^2}\frac{l\pi}{h},\enspace l= 2,4,\ldots,2N.\label{eq44}
\end{align}
After removing $B_{n}$'s and $C_{p}$'s, we can rewrite equations in a matrix form as
\begin{align}
\mathbf{M}_{1} \mathbf{U}^e + \mathbf{M}_{2}\mathbf{V}^o  &=\mathbf{0},\label{eq45}\\
\mathbf{M}_{3} \mathbf{U}^e + \mathbf{M}_{4}\mathbf{V}^o  &=\mathbf{0},\label{eq46}
\end{align}
where $\mathbf{M}_{i}$'s ($i=1,2,3,$ and $4$) are the $N\times N$ square matrices;
\begin{align}
M_{1,ij}&=\frac{4\beta}{h}\left(I_{2i-1,2j}\frac{2j}{\beta_{1,t}^2}+I_{2j,2i-1}\frac{2i-1}{\beta_{2,t}^2}\right),\label{eq47}\\
M_{2,ij}&=\delta_{ij}\frac{\omega \mu_{2} }{\beta_{2,t}^2}\alpha_{2,2i-1}
+\frac{16}{\pi^2}\frac{\omega \mu_{1}}{\beta_{1,t}^2} \left(\sum\limits_{p'=0}^{N}I_{2i-1,2p'}I_{2j-1,2p'}\alpha_{1,2p'}(\frac{-\delta_{2p'0}}{2}+1)\right),\label{eq48}\\
M_{3,ij}&=\delta_{ij}\frac{\omega \epsilon_{1}}{\beta_{1,t}^2}\alpha_{1,2i}
+\frac{16}{\pi^2}\frac{\omega \epsilon_{2}}{\beta_{2,t}^2} \sum\limits_{n'=1}^{N}I_{2i,2n'-1}I_{2j,2n'-1}\alpha_{2,2n'-1},\label{eq49}\\
M_{4,ij}&=\frac{4\beta}{h}\left(I_{2i,2j-1}\frac{2j-1}{\beta_{2,t}^2}+I_{2j-1,2i}\frac{2i}{\beta_{1,t}^2}\right),\label{eq50}
\end{align}
and $\mathbf{U}^e$ and $\mathbf{V}^o$ are the $N\times1$ column vectors;
\begin{align}
U_{j}^e&=A_{2j},\label{eq51}\\
V_{j}^o&=D_{2j-1}.\label{eq52}
\end{align}
If the matrix $\mathbf{M_2}$ is non-singular,  we can reduce the equations for $\mathbf{U}^e$;
\begin{equation}\label{eq53}
(\mathbf{M}_{3}-\mathbf{M}_{4}\mathbf{M}_{2}^{-1}\mathbf{M}_{1}) \mathbf{U}^e=\mathbf{0}.
\end{equation}
For the "non-trivial" solution to exist for the $\mathbf{U}^e$, 
\begin{equation}\label{eq54} 
\det[\mathbf{M}_{3}-\mathbf{M}_{4}\mathbf{M}_{2}^{-1}\mathbf{M}_{1}]=0.
\end{equation}
The latter set of equations has odd orders of $A_{m}$'s and $C_{p}$'s and even orders of $B_{n}$'s and $D_{q}$'s;
\begin{align}
   \sum\limits_{\text{odd}\; m}^{2N-1}A_{m}I_{ml}\frac{4}{\pi}&= B_{0}\delta_{l0}+B_{l},\enspace l = 0,2,\ldots,2N \label{eq55}\\
    \sum\limits_{\text{even}\; q}^{2N}D_{q}I_{ql}\frac{4}{\pi}&=C_{q},\enspace l = 1,3, \ldots,2N-1,\label{eq56}\\
   \sum\limits_{\text{odd}\; m}^{2N-1}A_{m}I_{lm}\frac{\beta}{\beta_{1,t}^2}\frac{m\pi}{h}\frac{4}{\pi}+
    \sum\limits_{\text{odd}\; p}^{2N-1}C_{p}I_{lp}\frac{\omega \mu_{1} }{\beta_{1,t}^2}\alpha_{1,p}\frac{4}{\pi}
   &=-B_{l}\frac{\beta}{\beta_{2,t}^2}\frac{l\pi}{h}-D_{l}\frac{\omega \mu_{2} }{\beta_{2,t}^2}\alpha_{2,l},\enspace l=2,4,\ldots,2N\label{eq57}\\
  \sum\limits_{\text{even}\; n=0}^{2N}B_{n}I_{ln}\frac{\omega\epsilon_{2}}{\beta_{2,t}^2}\alpha_{2,n}\frac{4}{\pi}+\sum\limits_{\text{even}\; q}^{2N}D_{q}I_{lq}\frac{\beta}{\beta_{2,t}^2}\frac{q\pi}{h}\frac{4}{\pi}
  &=-A_{l}\frac{\omega \epsilon_{1}}{\beta_{1,t}^2}\alpha_{1,l}-C_{l}\frac{\beta}{\beta_{1,t}^2}\frac{l\pi}{h},\enspace l= 1,3,\ldots,2N-1.\label{eq58}
\end{align}
 Similarly, after removing $B_{n}$'s and $C_{p}$'s, we can rewrite the equations in a matrix form as
\begin{align}
\mathbf{N}_{1} \mathbf{V}^{e} + \mathbf{N}_{2}\mathbf{U}^o  &=\mathbf{0},\label{eq59}\\
\mathbf{N}_{3} \mathbf{V}^{e} + \mathbf{N}_{4}\mathbf{U}^o  &=\mathbf{0},\label{eq60}
\end{align}
where $\mathbf{N}_{i}$'s ($i=1,2,3, \text{and} 4$) are $N\times N$  square matrices; 
\begin{align}
N_{1,ij}&=\frac{4\beta}{h}\left(I_{2i-1,2j}\frac{2j}{\beta_{2,t}^2}+I_{2j,2i-1}\frac{2j}{\beta_{1,t}^2}\right),\label{eq61}\\
N_{2,ij}&=\delta_{ij}\frac{\omega \epsilon_{1}}{\beta_{1,t}^2}\alpha_{1,2i-1}
+\frac{16}{\pi^2}\frac{\omega \epsilon_{2}}{\beta_{2,t}^2} \sum\limits_{n'=0}^{N}I_{2i-1,2n'}I_{2j-1,2n'}\alpha_{2,2n'}(\frac{-\delta_{2n'0}}{2}+1),\label{eq62}\\
N_{3,ij}&=\delta_{ij}\frac{\omega \mu_{2} }{\beta_{2,t}^2}\alpha_{2,2i}
+\frac{16}{\pi^2}\frac{\omega \mu_{1}}{\beta_{1,t}^2} \left(\sum\limits_{p'=1}^{N}I_{2i,2p'-1}I_{2j,2p'-1}\alpha_{1,2p'-1}\right),\label{eq63}\\
N_{4,ij}&=\frac{4\beta}{h}\left(I_{2i,2j-1}\frac{2j-1}{\beta_{1,t}^2}+I_{2j-1,2i}\frac{2i}{\beta_{2,t}^2}\right),\label{eq64}
\end{align}
and $\mathbf{U^o}$ and $\mathbf{V^e}$ are the $N\times 1$ column vectors;
\begin{align}
U_{j}^o&=A_{2j-1},\label{eq65}\\
V_{j}^e&=D_{2j}.\label{eq66}
\end{align}
If the matrix $\mathbf{N_2}$ is non-singular, we can reduce the equations for $\mathbf{V^{e}}$;
\begin{equation}
(\mathbf{N}_{3}-\mathbf{N}_{4}\mathbf{N}_{2}^{-1}\mathbf{N}_{1}) \mathbf{V}^e=\mathbf{0}.\label{eq67}
\end{equation}
For the "non-trivial" solution to exist for the $\mathbf{V^{e}}$, 
\begin{equation}\label{eq68} 
\det[\mathbf{N}_{3}-\mathbf{N}_{4}\mathbf{N}_{2}^{-1}\mathbf{N}_{1}]=0.
\end{equation}
The dispersion relation of the surface wave in this system is determined by either Eq. \eqref{eq54} or Eq. \eqref{eq68}, depending on the difference in refractive index $n$ between each region. For the existence of surface waves, the decay constant of all modes in each region from which the surface wave is composed should be real, satisfying the following condition; 
\begin{equation}\label{eq69}
    \beta^2-n_{j}^{2} k_0^{2}+k_{j,z}^2 > 0, \quad j=1,2.
\end{equation}
 When $n_1<n_2$, the surface waves follow the dispersion relation Eq. \eqref{eq54} while  Eq. \eqref{eq68} does not satisfy above condition. The dispersion relation is for the case where a TEM mode is incident on the PEC-PMC interface from the PEC region. Conversely, when $n_1>n_2$, the dispersion relation of the surface waves becomes Eq. \eqref{eq68} and Eq. \eqref{eq54} is discarded. The dispersion relation is for the case where a TEM mode is incident on the PEC-PMC interface from the PMC region. When $n_1=n_2$, there are no surface waves in this system. From the boundary conditions and Maxwell's equations, when the value of $\epsilon_1$ ($\mu_1$) is replaced by that of $\mu_2$ ($\epsilon_2$), the dispersion relation of the surface wave for the case where a TEM mode is incident on the PEC-PMC interface from the PMC region becomes identical to that for the case where a TEM mode is incident on the PEC-PMC interface from the PEC region.

\begin{figure}[htbp]
 \centering %
 \includegraphics[width=86mm]{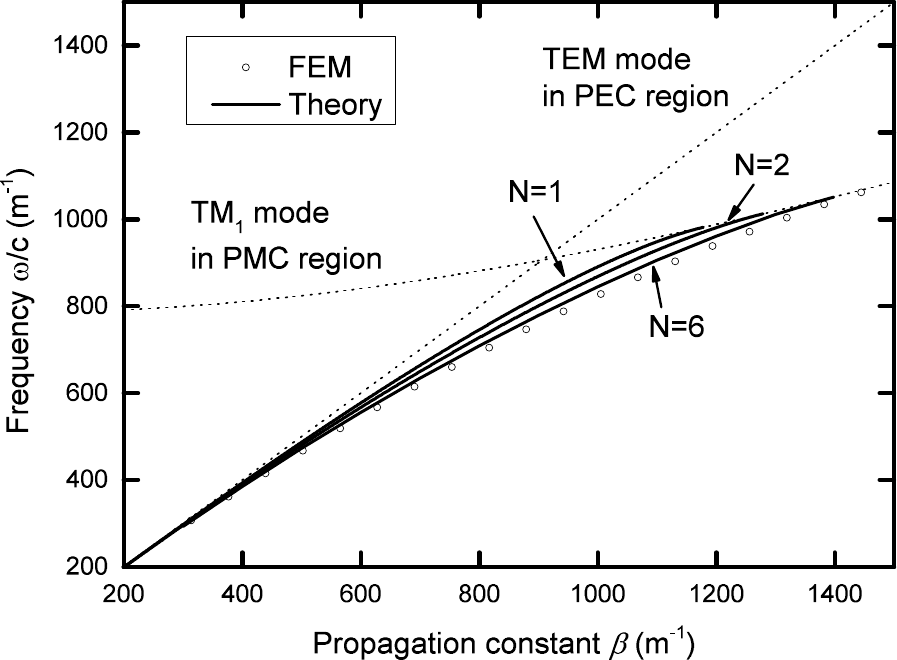}%
 \caption{The theoretical (solid line) and simulated (circles) dispersion relations of the surface waves at the PEC-PMC interface when $h=2$ mm, $\mu_1=1$, $\epsilon_1=1$, $\mu_2=1$, and $\epsilon_2=4$. A dotted straight line represents the light line of the TEM mode in the PEC parallel plate waveguide ($\omega=ck_0/n_1$), while a dotted curve denotes the dispersion curve of the TM$_{1}$ mode in the PMC parallel plate waveguide ($\omega=c\sqrt{k_0^2+(\pi/h)^2}/n_2$ \label{fig2} }
\end{figure}

\begin{figure}[htbp]
 \centering %
 \includegraphics[width=86mm]{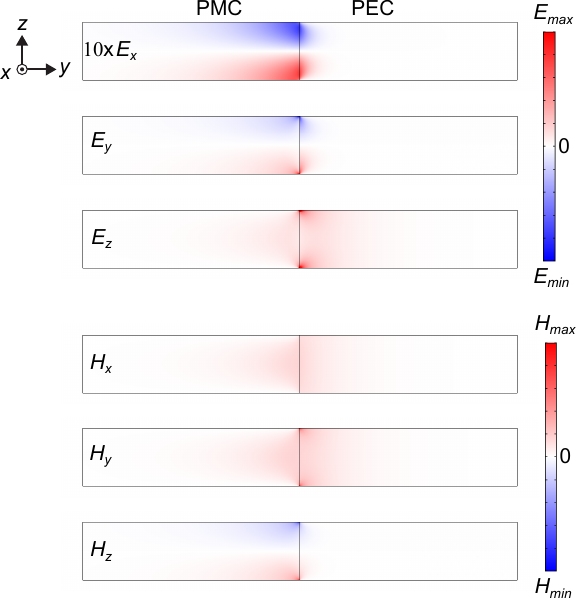}%
 \caption{The spatial distributions of the electric and magnetic field components of the surface wave in the $yz$ plane at a frequency of 46.36 GHz. \label{fig3} }
\end{figure}

\begin{figure}[htbp]
 \centering %
 \includegraphics[width=160mm]{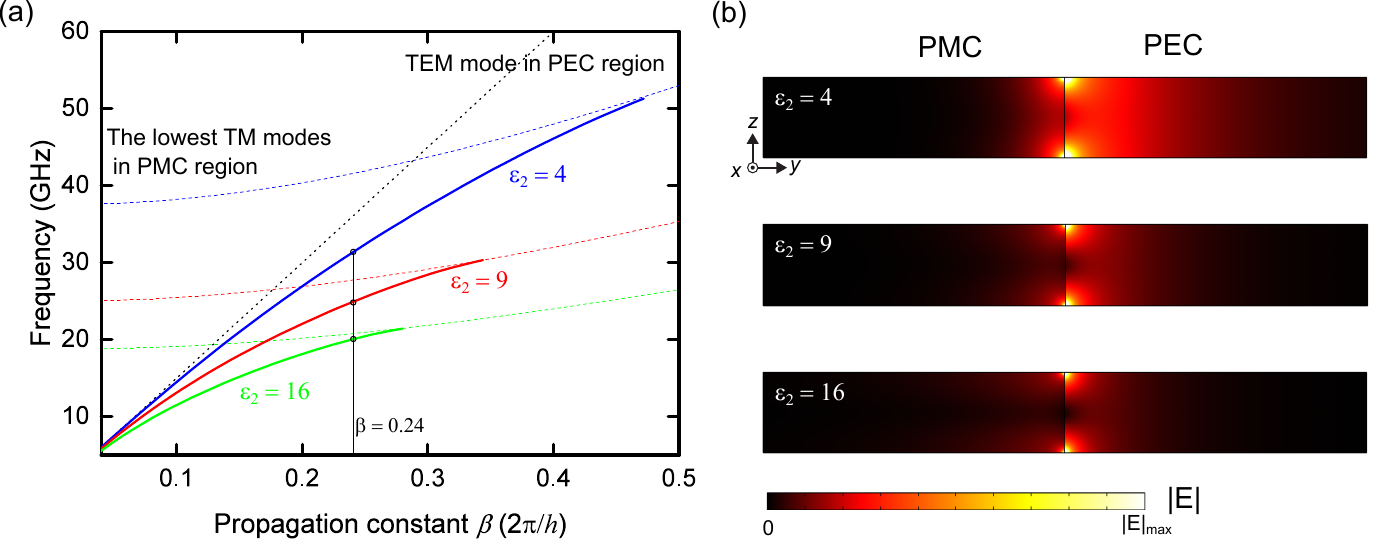}%
 \caption{(a) The simulated dispersion relations of the surface waves at the PEC-PMC interface for $\epsilon_2=$ 4, 9, and 16 when $h=2$ mm, $\mu_1=1$, $\epsilon_1=1$, and $\mu_2=1$. A dotted straight line represents the dispersion of the TEM mode in the PEC parallel plate waveguide ($\omega=ck_0/n_1$), while dotted curves denotes the dispersion curves of the TM$_{1}$ mode in the PMC parallel plate waveguide for $\epsilon_2 =$ 4, 9, and 16. (b) The spatial distributions of the electric field amplitude of the surface wave in the $yz$ plane at $\beta = 0.24(2\pi /h)$ for $\epsilon_2 =$ 4, 9, and 16. \label{fig_4a} }
\end{figure}

 Figure 2 shows dispersion relations of the surface waves for $N=1, 2,$ and 6 when $h= 2$ mm, $\mu_1=1$, $\epsilon_1=1$, $\mu_2=1$, and $\epsilon_2=4$. The analytic results are compared to those calculated by the finite element method (FEM). It is noted that the solution converges very quickly since the higher-order evanescent modes fast exponentially decay in $y$-direction. Even the result for just two modes ($N=2$) is very close to the FEM data.
 
A dotted straight line represents the light line of the TEM mode in the PEC parallel plate waveguide, $\omega=ck_0/n_1$, and a dotted curve shows the dispersion curve of the first-order TM (TM$_{1}$) mode in the PMC parallel plate waveguide, $\omega=c\sqrt{k_0^2+(\pi/h)^2}/n_2$. The dispersion curve of the surface waves lies below the light line and the dispersion curve of the TM$_{1}$ mode when they intersect ($n_{2} > n_{1}$). As the propagation constant of the surface wave increases, the dispersion curve of the surface wave gets closer to that of the TM$_{1}$ mode. Based on these properties, the dispersion relation of the surface wave can be easily sketched for cases with different $n_1, n_2$ and/or $h$. For example, when $n_2$ increases, the dispersion of the surface waves shifts away from the light line at the same frequency or propagation constant. This means that the surface mode becomes more tightly bound to the PEC-PMC interface as $n_2$increases (see Fig. 4). 

Figure 3 shows the spatial distributions of the electric and magnetic field components of the surface wave in the $yz$ plane at a frequency of 46.36 GHz. The field distributions show well the characteristics of the surface wave.  

Figure 4 represents the dispersion relations and the spatial distributions of the electric field amplitude of the surface waves at the PEC-PMC interface for $\epsilon_2=$ 4, 9, and 16. The confinement of fields is tighter by the larger dielectric constant in the PMC region, as shown in Fig. 4(b). 
 
It is valuable to discuss the differences between the surface wave and certain interface waves at the impedance interface formed by connecting inductive and capacitive surfaces\cite{lw1,lw2}, as well as the interface formed by connecting photonic crystals with different band geometric Zak phases\cite{sim1,sim2}. Firstly, line waves, which are interface waves at the impedance interface, cannot exist when it becomes a PEC-PMC interface; they are only loosely bound at the interface\cite{lw1}. Secondly, since the band geometric Zak phase is an inherent property of a periodic structure\cite{sim1}, the proposed PEC-PMC interface cannot support interface waves that occur at the interface of two periodic structures with different Zak phases. Therefore, the surface wave distinctly differs from the interface waves studied previously.

\subsection{Excitation of the surface wave by using prism coupling set-up}

\begin{figure}[htbp]
 \centering %
 \includegraphics[width=90mm]{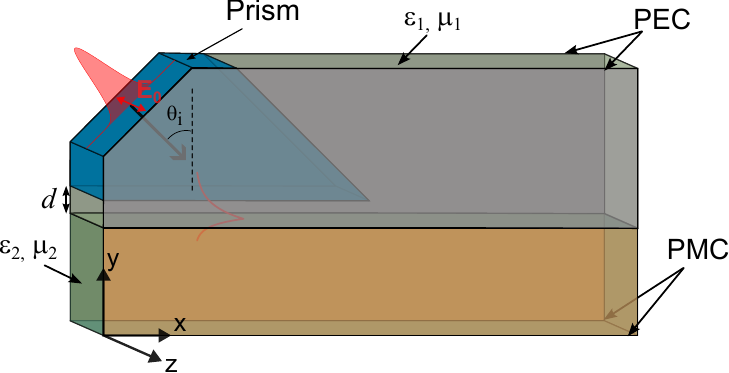}%
 \caption{Schematics of the prism coupling set-up based on the Otto configuration \label{fig_5a} }
\end{figure}

The propagation constant of the surface wave at a given frequency is greater than that of the TEM mode. As a result, the TEM mode from the PEC region cannot excite the surface wave at the PEC-PMC interface. Thus, the wave vector of the TEM mode from the PEC region must be increased to excite surface waves. Generally, in experiments, a prism coupling configuration has been employed to excite surface waves.\cite{prism1} An important concept underlying this setup involves the coupling of the surface wave with the evanescent wave, which arises due to attenuated total reflection at the base of a coupling prism. This occurs when a light beam is incident at an angle greater than the critical angle at the prism-air interface.

Figure 5 illuminates the prism coupling set-up based on the Otto configuration. A prism is placed in the PEC region. Here, $d$ denotes the distance between a prism and the PEC-PMC interface, and $\theta_i$ is an angle of incidence on the prism. TEM modes of the PEC waveguide are incident on the prism.

\begin{figure}[htbp]
\centering %
 \includegraphics[width=86mm]{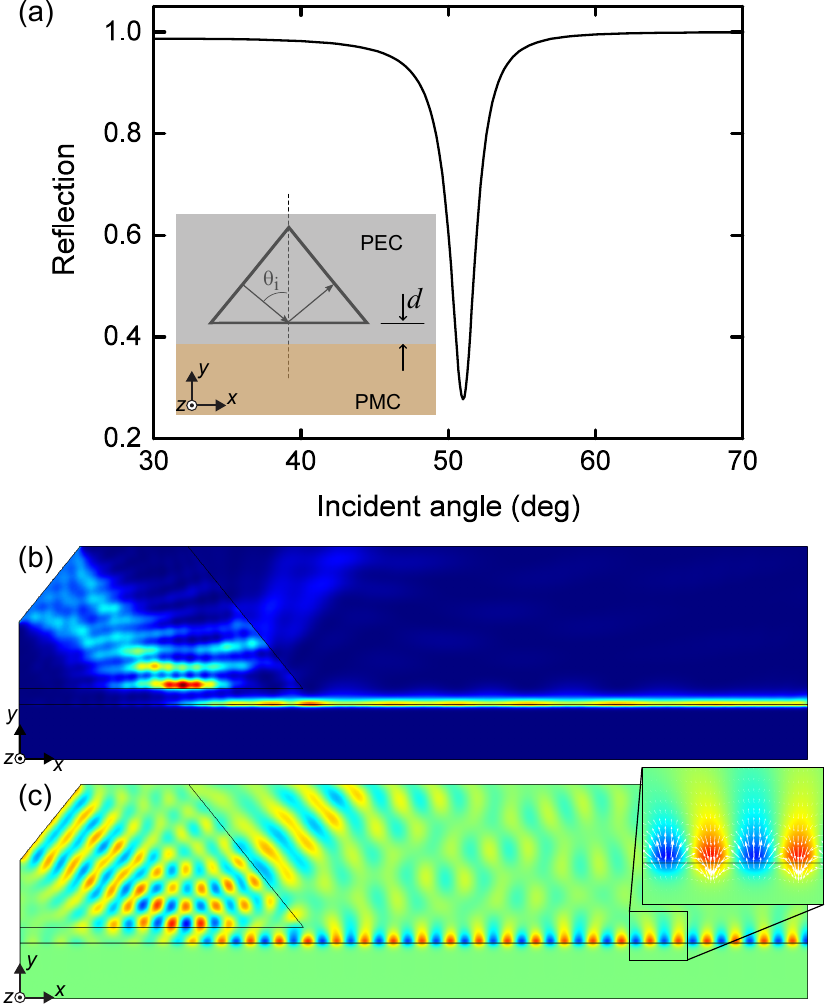}%
\caption{ (a) The reflection spectrum of the simulated prism coupling experiment for a frequency of 28 GHz when $h=2$ mm, $\mu_1=1$, $\epsilon_1=1$, $\mu_2=1$, $\epsilon_2=4+0.04i$, $d$ = 5 mm, and the index of refraction of the prism is 1.5. The inset shows a schematic of the prism coupling experiment configured using the Otto geometry. A prism is placed in the PEC region. $\theta_i$ denotes an angle of incidence on the prism and $d$ the distance between a prism and the PEC-PMC interface. (b) The spatial distribution of total electromagnetic energy density of the incident TEM wave and the excited surface wave and (c) the $E_z$ field component in the $xy$ plane at the height of $z = 1$ mm above the bottom plate when $\theta_i$ = 51 degrees. In the enlarged inset, white arrows display the directions of the magnetic field.\label{fig3_6a}}
\end{figure}

To verify the existence of the surface wave at the PEC-PMC interface, we performed a simulated prism coupling experiment configured using the Otto geometry,\cite{prism1} as illuminated in Fig. 5. A prism is placed in the PEC region. Here, $d$ denotes the distance between a prism and the PEC-PMC interface, and $\theta_i$ is an angle of incidence on the prism. An input wave can be coupled to the surface wave at the PEC-PMC interface when  $\theta_i > \theta_c$, where $\theta_c$ is the critical angle of incidence on the prism. In simulations, to clearly observe a resonance dip in the reflection spectrum due to the excitation of the surface wave, the imaginary part of $\epsilon_2$ was used, and $d$ was adjusted to a specific value.

Figure 6(a) presents the reflection spectrum of the simulated prism coupling experiment when $h=2$ mm, $\mu_1=1$, $\epsilon_1=1$, $\mu_2=1$, $\epsilon_2=4+0.04i$, $d$ = 5 mm, and the index of refraction of the prism is 1.5. The incident TEM wave has a frequency of 28 GHz. The spectrum clearly shows a resonance dip at an incident angle of 51 degrees attributed to the excitation of the surface wave. The excitation angle at this frequency closely aligns with the dispersion relation shown in Figure 2 (N=6). Figure 6(b) illustrates the spatial distribution of total electromagnetic energy density of the incident TEM wave and the excited surface wave in the $xy$ plane at the height of $z$=1 mm above the bottom plate. Figure 6(c) displays the $xy$ plane distributions of $E_z$ field component at the height of $z$=1 mm. The directions of the magnetic field are indicated by white arrows in the enlarged inset. In the PEC region, the directions of the magnetic field of the surface wave appear as though magnetic dipoles were being induced along the surface by the $y$ component of the magnetic field of the incident TEM wave, even though there are no actual induced magnetic dipoles on the PEC-PMC interface.

Comparing the characteristics of the surface wave at the PEC-PMC interface with those of a surface plasmon can provide valuable insights. In a surface plasmon, the electric field directions result from the collective oscillation of electrons on a metal surface, driven by the $x$ component of the electric field of an incident TM wave. The collective oscillation of electrons on a metal surface can be viewed as electric dipoles induced by the $x$ component of the electric field of an incident TM wave. At the PEC-PMC interface, based solely on the magnetic field directions of the surface wave, it can be inferred that they stem from virtual-induced magnetic dipoles. 

The results for the case where a TEM mode is incident on the PEC-PMC interface from the PMC region can be easily anticipated by the Maxwell's equations and boundary conditions. When the value of $\epsilon_1$ ($\mu_1$) is replaced by that of $\mu_2$ ($\epsilon_2$), the dispersion relation remains unchanged, and the electric and magnetic fields are interchanged. Therefore, the electric and magnetic field distributions of the surface wave in the PMC region resemble those of a surface plasmon, even though there are no free electrons on the surface, as shown in Fig. 7. 

 \begin{figure}[htbp]
 \centering %
 \includegraphics[width=160mm]{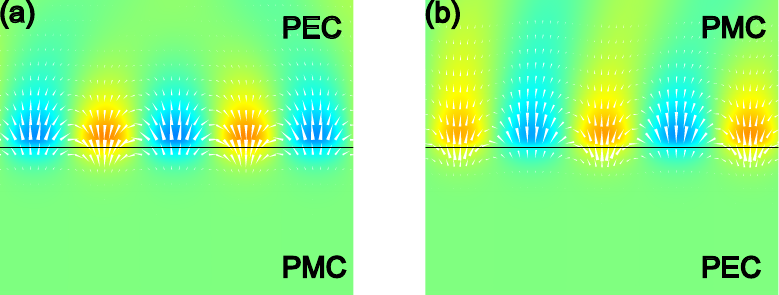}%
 \caption{(a) $E_z$ field component distribution in the $xy$ plane at a height of $z = 1$ mm above the bottom plate for prism coupling excitation from the PEC side with $\epsilon_1 = 1$, $\mu_1 = 1$, $\epsilon_2 = 4$, $\mu_2 = 1$, and $\theta_i = 48.8$ degrees. White arrows indicate the directions of the magnetic field  (b) $H_z$ field component distribution in the $xy$ plane at a height of $z = 1$ mm above the bottom plate for prism coupling excitation from the PMC side with $\epsilon_1 = 1$, $\mu_1 = 4$, $\epsilon_2 = 1$, $\mu_2 = 1$, and $\theta_i = 47$ degrees. White arrows display the directions of the electric field. \label{fig_7a} }
\end{figure}

\subsection{Existence of localized surface waves on an enclosed PEC-PMC surface }

\begin{figure}[htbp]
\centering  %
\includegraphics[width=80mm]{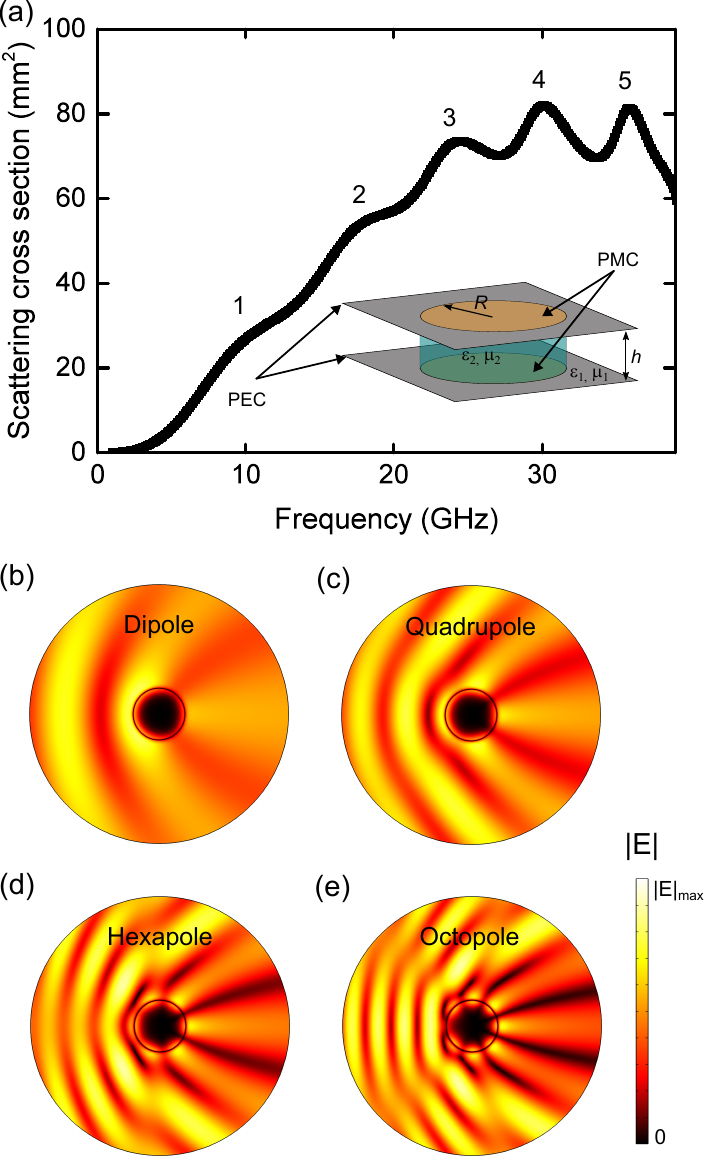}%
\caption{ (a) The simulated scattering cross-section (SCS) of a TEM mode from a cylinder with the PEC-PMC interface. The inset shows a schematic of the cylinder with a radius of $R$ containing a material with $\epsilon_{2}$ and $\mu_{2}$. (b) The spatial distribution of the electric field amplitude of the dipole resonance in the $xy$ plane at a height of $z = 1$ mm above the bottom plate and a frequency of 10.3 GHz when $h=2$ mm, $\mu_1=1$, $\epsilon_1=1$, $\mu_2=1$, $\epsilon_2=4$, and $R$ = 5 mm. (c) Quadrupole resonance at 17.4 GHz. (d) Hexapole resonance at 24.5 GHz. (e) Octopole resonance at 30.1 GHz. \label{fig_8a}}
\end{figure}

Localized surface plasmons (LSPs) in metal structures with enclosed surfaces have various applications, including near-field optics,\cite{lsp1} surface-enhanced spectroscopy,\cite{lsp2,lsp3,lsp4} plasmonic antennas,\cite{lsp5} and photovoltaics,\cite{lsp6,lsp7} due to the field enhancement associated with the LSP resonance. Therefore, studying the existence of localized surface waves on an enclosed PEC-PMC surface is valuable. For instance, we examine a circular cylinder with an enclosed PEC-PMC surface created by introducing identical circular PMC patches in the upper and bottom PEC parallel waveguide plates, as depicted in Figure 8(a). In simulations, $h=2$ mm, $\mu_1=1$, $\epsilon_1=1$, $\mu_2=1$, $\epsilon_2=4$, and $R$ = 5 mm.

The simulated scattering cross-section (SCS) of the TEM mode from the cylinder is shown in Figure 8(a). The SCS spectrum exhibits five distinct peaks, indicating resonant scattering. Figures 8(b)-(e) show the spatial distributions of the electric field amplitudes in the $xy$ plane at the height of $z = 1$ mm above the bottom plate for four different resonances: dipole resonance at 10.3 GHz, quadrupole resonance at 17.4 GHz, hexapole resonance at 24.5 GHz, and octopole resonance at 30.1 GHz. It is remarkable how the resonant response of the 2D cylinder closely resembles that of a Drude cylinder.\cite{lsp8} This observation strongly supports the presence of localized surface waves on the enclosed PEC-PMC surfaces. The cylinder placed in a PEC parallel plate waveguide without the PMC patches shows cavity modes more confined inside the cylinder, as illustrated in Fig. 9. The cylinder does not exhibit cavity modes that are more confined inside the cylinder, instead, surface modes are confined outside the cylinder, as represented in Fig. 9(b)-(e). 

\begin{figure}[htbp]
 \centering %
 \includegraphics[width=90mm]{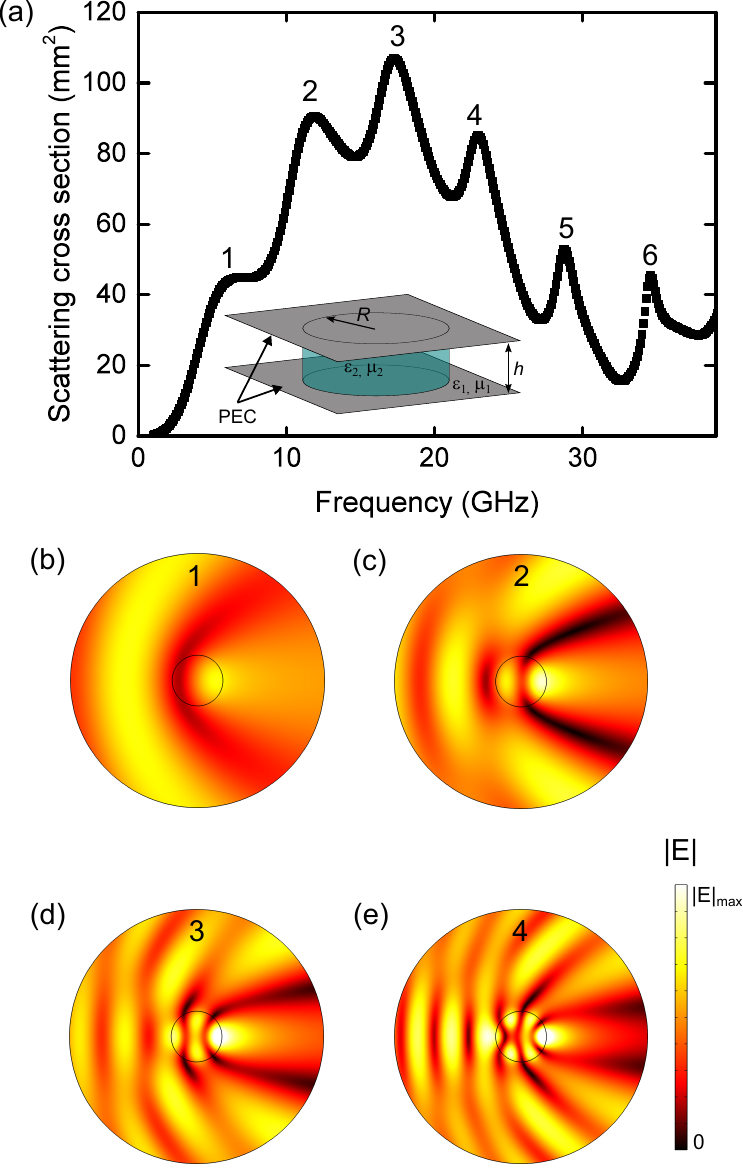}%
 \caption{ (a) The simulated scattering cross-section of TEM modes from a cylinder within the PEC parallel waveguide. The inset shows a schematic of the dielectric cylinder with $\epsilon_{2}$, $\mu_{2}$, and a radius of $R$. (b) The spatial distribution of the electric field amplitude of the first resonance in the $xy$ plane at a height of $z = 1$ mm above the bottom plate and a frequency of 6.9 GHz when $h=2$ mm, $\mu_1=1$, $\epsilon_1=1$, $\mu_2=1$, $\epsilon_2=4$, and $R$ = 5 mm. (c) The second resonance at 11.9 GHz. (d) The third resonance at 17.3 GHz. (e) The fourth resonance at 22.9 GHz. \label{fig_9a} }
\end{figure}

Observing the excitation of surface waves at the interface between a PEC and PMC parallel plate waveguides experimentally is challenging because PMC is an idealized material that has a zero magnetic field on its surface. Although physically perfect PMCs do not exist, since the concept of an ideal PMC is extensively employed in antenna design, an artificial structure known as an artificial magnetic conductor (AMC) has been developed, capable of replicating the functionality of a PMC within a narrow frequency band.\cite{amc0,amc1}  For example, AMC substrates consist of a frequency-selective surface that blocks or reflects waves in specific frequency bands.\cite{amc2,amc3} These substrates are used to enhance antenna efficiency and control frequency selectivity. Recently, it has been reported that AMCs based on metamaterials artificially designed materials with unique geometrical structures can serve as core components of AMCs.\cite{amc4,amc5} Therefore, it could potentially enable experimental verification of the proposed surface wave if broadband AMCs are developed.

\subsection{Demonstrating the existence and excitation of surface waves using a practical structure}

\begin{figure}[htbp]
\centering %
\includegraphics[width=76mm]{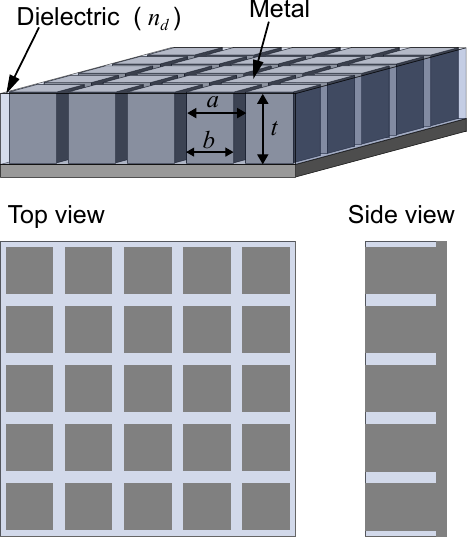}%
 \caption{ Schematics of a proposed artificial magnetic conductor (AMC), which is composed of a two-dimensional array of square metal rods on a flat metal surface within a dielectric background ($n_d$). $a$ is the period of the array. $b$ and $t$ denote the width and height of the rods, respectively. The top and side views are also shown at the bottom. \label{fig_01a} }
\end{figure}

\begin{figure}[htbp]
\centering %
\includegraphics[width=160mm]{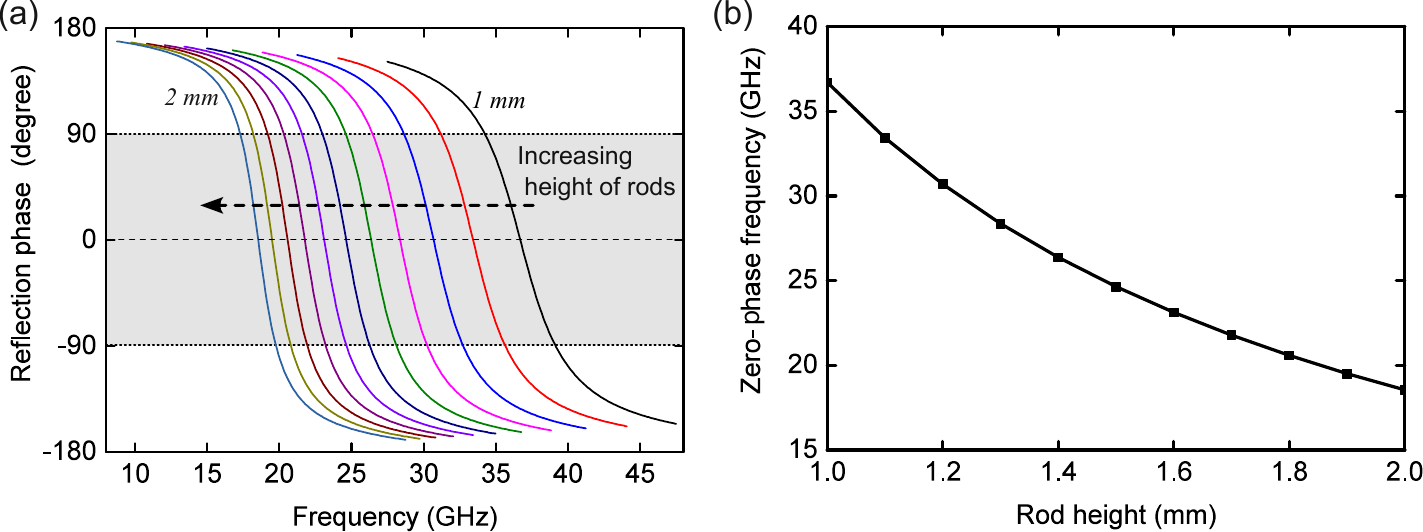}%
 \caption{ (a) Reflection phase changes of normally incident electromagnetic waves on the AMC with $a$ = 1.0 mm, $b$ = 0.8$a$ and $n_d = 2$ when $t$ varies from 1 to 2 mm. The artificial structure acts as an AMC when the phase change is between 90 and -90 degrees, as indicated in gray.  (b) Zero-phase frequency as a function of $t$ .   \label{fig_11a} }
\end{figure}

\begin{figure}[htbp]
 \centering %
 \includegraphics[width=160mm]{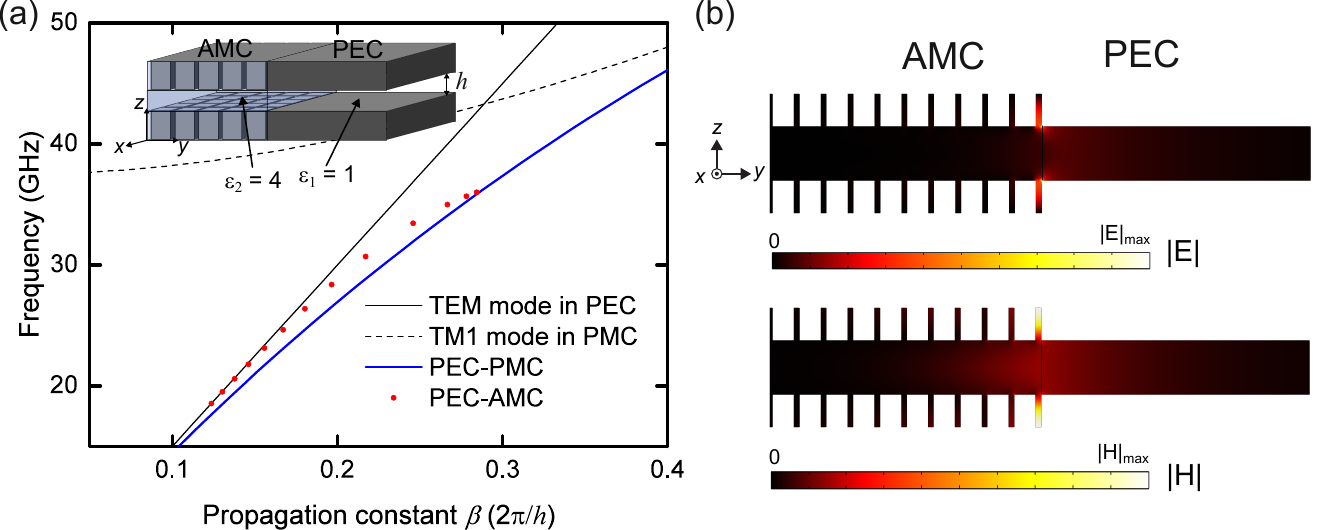}%
\caption{ (a) Dispersion relation of the surface wave at the interface of the PEC-AMC parallel waveguide structure as $t$ varies from 1 to 2 mm, with $a = $ 1 mm, $b =0.8a$, $n_d = 2$, $h=$ 2 mm, $\epsilon_{1}=1.0$ and $\epsilon_{2}=4.0$. A schematic of the PEC-AMC parallel waveguide structure is shown in the inset. (b) Intensity distributions of the electric fields (upper) and magnetic fields (lower) of the surface wave for the PEC-AMC case at the zero-phase frequency of 30.687 GHz  with $t$ = 1.2 mm.  \label{fig_12a }}
\end{figure}

\begin{figure}[htbp]
 \centering %
 \includegraphics[width=160mm]{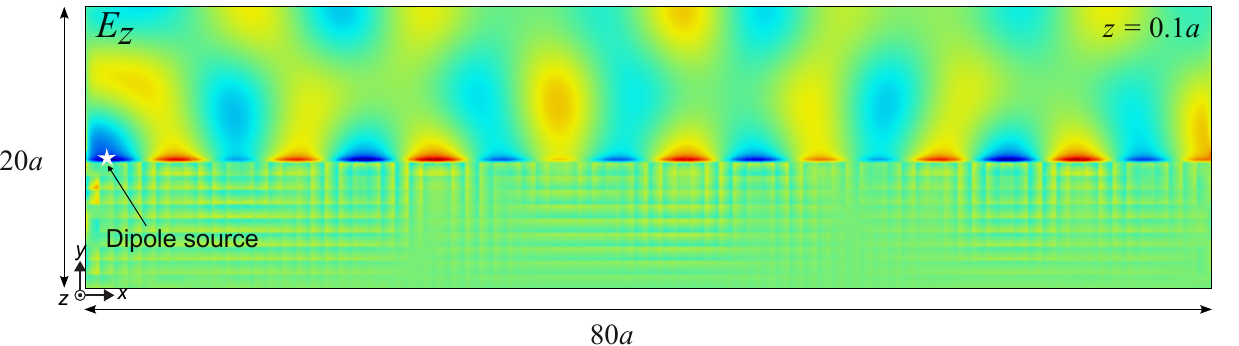}%
\caption{Spatial distribution of the $E_z$ field in the $xy$ plane at a height of $z = 0.1a$ above the bottom plate for a frequency, 30.687 GHz, which corresponds to the zero-phase frequency of the AMC with $t = $1.2 mm, $a = $1 mm, $b =0.8a$, $h=$ 2 mm, and $n_d = 2$ when a dipole source with the zero-phase frequency is applied near the PEC-AMC interface. The PEC-AMC parallel waveguide structure has dimensions of $80a \times 20a$ and is surrounded by perfectly matched layers to minimize edge reflections.  \label{fig_13a }}
\end{figure}

Figure 10 shows the schematics of a proposed artificial magnetic conductor (AMC), which is composed of a two-dimensional array of square metal rods on a flat metal surface within a dielectric background ($n_d$). $a$ is the period of the array. $b$ and $t$ denote the width and height of the rods, respectively. The top and side views are also shown at the bottom.

The reflection phase changes of normally incident electromagnetic waves on the AMC are shown in Figure 11(a). $t$ varies from 1 to 2 mm with $a$ = 1.0 mm, $b$ = 0.8$a$ and $n_d = 2$. In general, the artificial structure can act as a magnetic conductor in a frequency range where the phase change is between 90 and -90 degrees, as indicated in gray. Figure 11(b) shows the zero-phase frequency as a function of $t$. The existence of the surface wave at the interface of the PEC-AMC parallel waveguide structure was examined at the zero-phase frequency.

Figure 12(a) shows the dispersion relation of the surface wave at the interface of the PEC-AMC parallel waveguide structure as $t$ varies from 1 to 2 mm, with $a$ = 1 mm, $b$ = 0.8$a$, $h=$ 2 mm, and $n_d = 2$, $\epsilon_{1}=1.0$ and $\epsilon_{2}=4.0$. A schematic of the PEC-AMC parallel waveguide structure is shown in the inset. Comparing the dispersion relations for the PEC-PMC and PEC-AMC cases reveals that the surface wave confinement at the interface is more pronounced in the PEC-PMC case. This is because the dispersion curve for the PEC-PMC case is situated further from the light line compared to the PEC-AMC case. Figure 12(b) illustrates the intensity distributions of the electric fields (upper) and magnetic fields (lower) of the surface wave for the PEC-AMC case at 30.687 GHz, which is the zero-phase frequency of the AMC with $t$ =1.2 mm.

Figure 13 shows the spatial distribution of the $E_z$ field in the $xy$ plane at a height of $z = 0.1a$ above the bottom plate for a frequency, 30.687 GHz, which corresponds to the zero-phase frequency of the AMC with $t =$ 1.2 mm, $a =$ 1 mm, $b=0.8a$, $h=$ 2 mm and $n_d = 2$ when a dipole source with the zero-phase frequency is applied near the PEC-AMC interface. The PEC-AMC parallel waveguide structure in the simulation has dimensions of $80a \times 20a$ and is surrounded by perfectly matched layers to minimize edge reflections. Using prism coupling to excite the surface wave at the PEC-AMC interface would require a much larger computational space than $80a \times 20a$, which is beyond our computational capacity. Instead, a dipole source effectively excites the surface wave at the PEC-AMC interface, even though this method of coupling is not resonant like prism coupling. In this setup, the near-field wave near the dipole source partially couples to the surface wave, while the far-field wave propagates through the PEC waveguide region. The simulation validates the existence and excitation of the surface wave on the PEC-AMC interface.


\section{Conclusion}
In conclusion, we have demonstrated the existence of surface waves at the interface between a PEC and a PMC parallel plate waveguides containing materials with positive permittivity or permeability. This challenges the conventional understanding that surface waves require materials with negative permittivity or permeability. The results of the simulated prism coupling experiment provide strong evidence of these surface waves. Additionally, the resonant response of the 2D cylinder with the enclosed PEC-PMC surface closely resembles that of a Drude cylinder. The development of artificial structures like broadband AMCs might offer promise for experimental verification. Our discovery broadens the understanding of the conditions for generating electromagnetic surface waves.


\begin{thebibliography}{29}%
\makeatletter
\providecommand \@ifxundefined [1]{%
 \@ifx{#1\undefined}
}%
\providecommand \@ifnum [1]{%
 \ifnum #1\expandafter \@firstoftwo
 \else \expandafter \@secondoftwo
 \fi
}%
\providecommand \@ifx [1]{%
 \ifx #1\expandafter \@firstoftwo
 \else \expandafter \@secondoftwo
 \fi
}%
\providecommand \natexlab [1]{#1}%
\providecommand \enquote  [1]{``#1''}%
\providecommand \bibnamefont  [1]{#1}%
\providecommand \bibfnamefont [1]{#1}%
\providecommand \citenamefont [1]{#1}%
\providecommand \href@noop [0]{\@secondoftwo}%
\providecommand \href [0]{\begingroup \@sanitize@url \@href}%
\providecommand \@href[1]{\@@startlink{#1}\@@href}%
\providecommand \@@href[1]{\endgroup#1\@@endlink}%
\providecommand \@sanitize@url [0]{\catcode `\\12\catcode `\$12\catcode `\&12\catcode `\#12\catcode `\^12\catcode `\_12\catcode `\%12\relax}%
\providecommand \@@startlink[1]{}%
\providecommand \@@endlink[0]{}%
\providecommand \url  [0]{\begingroup\@sanitize@url \@url }%
\providecommand \@url [1]{\endgroup\@href {#1}{\urlprefix }}%
\providecommand \urlprefix  [0]{URL }%
\providecommand \Eprint [0]{\href }%
\providecommand \doibase [0]{https://doi.org/}%
\providecommand \selectlanguage [0]{\@gobble}%
\providecommand \bibinfo  [0]{\@secondoftwo}%
\providecommand \bibfield  [0]{\@secondoftwo}%
\providecommand \translation [1]{[#1]}%
\providecommand \BibitemOpen [0]{}%
\providecommand \bibitemStop [0]{}%
\providecommand \bibitemNoStop [0]{.\EOS\space}%
\providecommand \EOS [0]{\spacefactor3000\relax}%
\providecommand \BibitemShut  [1]{\csname bibitem#1\endcsname}%
\let\auto@bib@innerbib\@empty
\bibitem [{\citenamefont {Agranovich}\ and\ \citenamefont {Mills}(1982)}]{sw1}%
  \BibitemOpen
  \bibfield  {author} {\bibinfo {author} {\bibfnamefont {V.~M.}\ \bibnamefont {Agranovich}}\ and\ \bibinfo {author} {\bibfnamefont {D.~L.}\ \bibnamefont {Mills}},\ }\href@noop {} {\emph {\bibinfo {title} {Surface Polaritons: Electromagnetic Waves at Surfaces and Interfaces}}}\ (\bibinfo  {publisher} {North-Holland Publishing Company},\ \bibinfo {year} {1982})\BibitemShut {NoStop}%
\bibitem [{\citenamefont {Brongersma}\ and\ \citenamefont {Kik}(2007)}]{sw3}%
  \BibitemOpen
  \bibfield  {author} {\bibinfo {author} {\bibfnamefont {M.~L.}\ \bibnamefont {Brongersma}}\ and\ \bibinfo {author} {\bibfnamefont {P.~G.}\ \bibnamefont {Kik}},\ }\href@noop {} {\emph {\bibinfo {title} {Surface Plasmon Nanophotonics}}}\ (\bibinfo  {publisher} {Springer Series in Optical Sciences},\ \bibinfo {year} {2007})\BibitemShut {NoStop}%
\bibitem [{\citenamefont {Yu}\ \emph {et~al.}(2019)\citenamefont {Yu}, \citenamefont {Peng}, \citenamefont {Yang},\ and\ \citenamefont {Li}}]{sw4}%
  \BibitemOpen
  \bibfield  {author} {\bibinfo {author} {\bibfnamefont {H.}~\bibnamefont {Yu}}, \bibinfo {author} {\bibfnamefont {Y.}~\bibnamefont {Peng}}, \bibinfo {author} {\bibfnamefont {Y.}~\bibnamefont {Yang}},\ and\ \bibinfo {author} {\bibfnamefont {Z.-Y.}\ \bibnamefont {Li}},\ }\bibfield  {title} {\bibinfo {title} {Plasmon-enhanced light–matter interactions and applications.},\ }\href@noop {} {\bibfield  {journal} {\bibinfo  {journal} {npj Comput Mater.}\ }\textbf {\bibinfo {volume} {5}},\ \bibinfo {pages} {45} (\bibinfo {year} {2019})}\BibitemShut {NoStop}%
\bibitem [{\citenamefont {Maier}(2007)}]{sw2}%
  \BibitemOpen
  \bibfield  {author} {\bibinfo {author} {\bibfnamefont {S.}~\bibnamefont {Maier}},\ }\href@noop {} {\emph {\bibinfo {title} {Plasmonics: Fundamentals and Applications}}}\ (\bibinfo  {publisher} {Springer, New York},\ \bibinfo {year} {2007})\BibitemShut {NoStop}%
\bibitem [{\citenamefont {Volakis}\ and\ \citenamefont {L.}(2019)}]{sw5}%
  \BibitemOpen
  \bibfield  {author} {\bibinfo {author} {\bibnamefont {Volakis}}\ and\ \bibinfo {author} {\bibfnamefont {J.}~\bibnamefont {L.}},\ }\href@noop {} {\emph {\bibinfo {title} {Antenna Engineering Handbook, 5th ed, Ch. 10}}}\ (\bibinfo  {publisher} {New York: McGraw-Hill},\ \bibinfo {year} {2019})\BibitemShut {NoStop}%
\bibitem [{\citenamefont {Li}\ \emph {et~al.}(2008)\citenamefont {Li}, \citenamefont {Boone}, \citenamefont {Bozzi}, \citenamefont {Perregrini},\ and\ \citenamefont {Wu}}]{pmc1}%
  \BibitemOpen
  \bibfield  {author} {\bibinfo {author} {\bibfnamefont {D.~C.}\ \bibnamefont {Li}}, \bibinfo {author} {\bibfnamefont {F.}~\bibnamefont {Boone}}, \bibinfo {author} {\bibfnamefont {M.}~\bibnamefont {Bozzi}}, \bibinfo {author} {\bibfnamefont {L.}~\bibnamefont {Perregrini}},\ and\ \bibinfo {author} {\bibfnamefont {K.}~\bibnamefont {Wu}},\ }\bibfield  {title} {\bibinfo {title} {Concept of virtual electric/magnetic walls and its realization with artificial magnetic conductor technique},\ }\href@noop {} {\bibfield  {journal} {\bibinfo  {journal} {IEEE Microwave and Wireless Components Letters}\ }\textbf {\bibinfo {volume} {18}},\ \bibinfo {pages} {743} (\bibinfo {year} {2008})}\BibitemShut {NoStop}%
\bibitem [{\citenamefont {Kim}\ and\ \citenamefont {Kee}(2019)}]{pmc3}%
  \BibitemOpen
  \bibfield  {author} {\bibinfo {author} {\bibfnamefont {S.-H.}\ \bibnamefont {Kim}}\ and\ \bibinfo {author} {\bibfnamefont {C.-S.}\ \bibnamefont {Kee}},\ }\bibfield  {title} {\bibinfo {title} {Electromagnetic shielding via a virtual pillar with a magnetic wall},\ }\href@noop {} {\bibfield  {journal} {\bibinfo  {journal} {Results in Physics}\ }\textbf {\bibinfo {volume} {14}},\ \bibinfo {pages} {102462} (\bibinfo {year} {2019})}\BibitemShut {NoStop}%
\bibitem [{\citenamefont {Kim}\ \emph {et~al.}(2016)\citenamefont {Kim}, \citenamefont {Kim},\ and\ \citenamefont {Kee}}]{pmc2}%
  \BibitemOpen
  \bibfield  {author} {\bibinfo {author} {\bibfnamefont {S.-H.}\ \bibnamefont {Kim}}, \bibinfo {author} {\bibfnamefont {S.}~\bibnamefont {Kim}},\ and\ \bibinfo {author} {\bibfnamefont {C.-S.}\ \bibnamefont {Kee}},\ }\bibfield  {title} {\bibinfo {title} {Photonic crystals composed of virtual pillars with magnetic walls: Photonic band gaps and double dirac cones},\ }\href@noop {} {\bibfield  {journal} {\bibinfo  {journal} {Phys. Rev. B}\ }\textbf {\bibinfo {volume} {94}},\ \bibinfo {pages} {085118} (\bibinfo {year} {2016})}\BibitemShut {NoStop}%
\bibitem [{sm1()}]{sm1}%
  \BibitemOpen
  \href@noop {} {\bibinfo  {journal} {See the Supplemental Information for the mathematical derivation of the master matrix equation used to determine the dispersion relation of the surface wave at the PEC-PMC interface through a modal expansion of EM fields.}\ }\BibitemShut {NoStop}%
\bibitem [{\citenamefont {Bisharat}\ and\ \citenamefont {Sievenpiper}(2017)}]{lw1}%
  \BibitemOpen
\bibfield  {journal} {  }\bibfield  {author} {\bibinfo {author} {\bibfnamefont {D.~J.}\ \bibnamefont {Bisharat}}\ and\ \bibinfo {author} {\bibfnamefont {D.~F.}\ \bibnamefont {Sievenpiper}},\ }\bibfield  {title} {\bibinfo {title} {Guiding waves along an infinitesimal line between impedance surfaces},\ }\href {https://doi.org/10.1103/PhysRevLett.119.106802} {\bibfield  {journal} {\bibinfo  {journal} {Phys. Rev. Lett.}\ }\textbf {\bibinfo {volume} {119}},\ \bibinfo {pages} {106802} (\bibinfo {year} {2017})}\BibitemShut {NoStop}%
\bibitem [{\citenamefont {Xu}\ \emph {et~al.}(2021)\citenamefont {Xu}, \citenamefont {Chang}, \citenamefont {Fang}, \citenamefont {Zhang}, \citenamefont {Davis}, \citenamefont {Sievenpiper},\ and\ \citenamefont {Cui}}]{lw2}%
  \BibitemOpen
  \bibfield  {author} {\bibinfo {author} {\bibfnamefont {Z.}~\bibnamefont {Xu}}, \bibinfo {author} {\bibfnamefont {J.}~\bibnamefont {Chang}}, \bibinfo {author} {\bibfnamefont {S.}~\bibnamefont {Fang}}, \bibinfo {author} {\bibfnamefont {Q.}~\bibnamefont {Zhang}}, \bibinfo {author} {\bibfnamefont {R.~J.}\ \bibnamefont {Davis}}, \bibinfo {author} {\bibfnamefont {D.}~\bibnamefont {Sievenpiper}},\ and\ \bibinfo {author} {\bibfnamefont {T.~J.}\ \bibnamefont {Cui}},\ }\bibfield  {title} {\bibinfo {title} {Line waves existing at junctions of dual-impedance metasurfaces},\ }\href {https://doi.org/10.1021/acsphotonics.1c00344} {\bibfield  {journal} {\bibinfo  {journal} {ACS Photonics}\ }\textbf {\bibinfo {volume} {8}},\ \bibinfo {pages} {2285} (\bibinfo {year} {2021})}\BibitemShut {NoStop}%
\bibitem [{\citenamefont {Xiao}\ \emph {et~al.}(2014)\citenamefont {Xiao}, \citenamefont {Zhang},\ and\ \citenamefont {Chan}}]{sim1}%
  \BibitemOpen
  \bibfield  {author} {\bibinfo {author} {\bibfnamefont {M.}~\bibnamefont {Xiao}}, \bibinfo {author} {\bibfnamefont {Z.~Q.}\ \bibnamefont {Zhang}},\ and\ \bibinfo {author} {\bibfnamefont {C.~T.}\ \bibnamefont {Chan}},\ }\bibfield  {title} {\bibinfo {title} {Surface impedance and bulk band geometric phases in one-dimensional systems},\ }\href {https://doi.org/10.1103/PhysRevX.4.021017} {\bibfield  {journal} {\bibinfo  {journal} {Phys. Rev. X}\ }\textbf {\bibinfo {volume} {4}},\ \bibinfo {pages} {021017} (\bibinfo {year} {2014})}\BibitemShut {NoStop}%
\bibitem [{\citenamefont {Xiao}\ \emph {et~al.}(2016)\citenamefont {Xiao}, \citenamefont {Huang}, \citenamefont {Fang},\ and\ \citenamefont {Chan}}]{sim2}%
  \BibitemOpen
  \bibfield  {author} {\bibinfo {author} {\bibfnamefont {M.}~\bibnamefont {Xiao}}, \bibinfo {author} {\bibfnamefont {X.}~\bibnamefont {Huang}}, \bibinfo {author} {\bibfnamefont {A.}~\bibnamefont {Fang}},\ and\ \bibinfo {author} {\bibfnamefont {C.~T.}\ \bibnamefont {Chan}},\ }\bibfield  {title} {\bibinfo {title} {Effective impedance for predicting the existence of surface states},\ }\href {https://doi.org/10.1103/PhysRevB.93.125118} {\bibfield  {journal} {\bibinfo  {journal} {Phys. Rev. B}\ }\textbf {\bibinfo {volume} {93}},\ \bibinfo {pages} {125118} (\bibinfo {year} {2016})}\BibitemShut {NoStop}%
\bibitem [{\citenamefont {Sambles}\ \emph {et~al.}(1991)\citenamefont {Sambles}, \citenamefont {Bradbery},\ and\ \citenamefont {Yang}}]{prism1}%
  \BibitemOpen
  \bibfield  {author} {\bibinfo {author} {\bibfnamefont {J.~R.}\ \bibnamefont {Sambles}}, \bibinfo {author} {\bibfnamefont {G.~W.}\ \bibnamefont {Bradbery}},\ and\ \bibinfo {author} {\bibfnamefont {F.}~\bibnamefont {Yang}},\ }\bibfield  {title} {\bibinfo {title} {Optical excitation of surface plasmons: an introduction},\ }\href@noop {} {\bibfield  {journal} {\bibinfo  {journal} {Contemporary Physics}\ }\textbf {\bibinfo {volume} {32}},\ \bibinfo {pages} {173} (\bibinfo {year} {1991})}\BibitemShut {NoStop}%
\bibitem [{sm2()}]{sm2}%
  \BibitemOpen
  \href@noop {} {\bibinfo  {journal} {See the Supplemental Information for the spatial distributions of the electric and magnetic field components of the surface wave for the case where a TEM mode is incident on the PEC-PMC interface from the PMC region.}\ }\BibitemShut {NoStop}%
\bibitem [{\citenamefont {Novotny}\ and\ \citenamefont {Hecht}(2006)}]{lsp1}%
  \BibitemOpen
\bibfield  {journal} {  }\bibfield  {author} {\bibinfo {author} {\bibfnamefont {L.}~\bibnamefont {Novotny}}\ and\ \bibinfo {author} {\bibfnamefont {B.}~\bibnamefont {Hecht}},\ }\href@noop {} {\emph {\bibinfo {title} {Principles of Nano-Optics}}}\ (\bibinfo  {publisher} {Cambridge University Press, Cambridge},\ \bibinfo {year} {2006})\BibitemShut {NoStop}%
\bibitem [{\citenamefont {Garc\'{\i}a-Vidal}\ and\ \citenamefont {Pendry}(1996)}]{lsp2}%
  \BibitemOpen
  \bibfield  {author} {\bibinfo {author} {\bibfnamefont {F.~J.}\ \bibnamefont {Garc\'{\i}a-Vidal}}\ and\ \bibinfo {author} {\bibfnamefont {J.~B.}\ \bibnamefont {Pendry}},\ }\bibfield  {title} {\bibinfo {title} {Collective theory for surface enhanced raman scattering},\ }\href {https://doi.org/10.1103/PhysRevLett.77.1163} {\bibfield  {journal} {\bibinfo  {journal} {Phys. Rev. Lett.}\ }\textbf {\bibinfo {volume} {77}},\ \bibinfo {pages} {1163} (\bibinfo {year} {1996})}\BibitemShut {NoStop}%
\bibitem [{\citenamefont {Prodan}\ \emph {et~al.}(2003)\citenamefont {Prodan}, \citenamefont {Radloff}, \citenamefont {Halas},\ and\ \citenamefont {Nordlander}}]{lsp3}%
  \BibitemOpen
  \bibfield  {author} {\bibinfo {author} {\bibfnamefont {E.}~\bibnamefont {Prodan}}, \bibinfo {author} {\bibfnamefont {C.}~\bibnamefont {Radloff}}, \bibinfo {author} {\bibfnamefont {N.~J.}\ \bibnamefont {Halas}},\ and\ \bibinfo {author} {\bibfnamefont {P.}~\bibnamefont {Nordlander}},\ }\bibfield  {title} {\bibinfo {title} {A hybridization model for the plasmon response of complex nanostructures},\ }\href@noop {} {\bibfield  {journal} {\bibinfo  {journal} {Science}\ }\textbf {\bibinfo {volume} {302}},\ \bibinfo {pages} {419} (\bibinfo {year} {2003})}\BibitemShut {NoStop}%
\bibitem [{\citenamefont {Anker}\ \emph {et~al.}(2008)\citenamefont {Anker}, \citenamefont {Hall}, \citenamefont {Lyandres}, \citenamefont {Shah}, \citenamefont {Zhao},\ and\ \citenamefont {Duyne}}]{lsp4}%
  \BibitemOpen
  \bibfield  {author} {\bibinfo {author} {\bibfnamefont {J.~N.}\ \bibnamefont {Anker}}, \bibinfo {author} {\bibfnamefont {W.~P.}\ \bibnamefont {Hall}}, \bibinfo {author} {\bibfnamefont {O.}~\bibnamefont {Lyandres}}, \bibinfo {author} {\bibfnamefont {N.}~\bibnamefont {Shah}}, \bibinfo {author} {\bibfnamefont {J.}~\bibnamefont {Zhao}},\ and\ \bibinfo {author} {\bibfnamefont {R.~P.~V.}\ \bibnamefont {Duyne}},\ }\bibfield  {title} {\bibinfo {title} {Biosensing with plasmonic nanosensors},\ }\href@noop {} {\bibfield  {journal} {\bibinfo  {journal} {Nature Mater.}\ }\textbf {\bibinfo {volume} {7}},\ \bibinfo {pages} {442} (\bibinfo {year} {2008})}\BibitemShut {NoStop}%
\bibitem [{\citenamefont {Schnell}\ \emph {et~al.}(2009)\citenamefont {Schnell}, \citenamefont {Garcı´a-Etxarri}, \citenamefont {andK. Crozier}, \citenamefont {Aizpurua},\ and\ \citenamefont {Hillenbran}}]{lsp5}%
  \BibitemOpen
  \bibfield  {author} {\bibinfo {author} {\bibfnamefont {M.}~\bibnamefont {Schnell}}, \bibinfo {author} {\bibfnamefont {A.}~\bibnamefont {Garcı´a-Etxarri}}, \bibinfo {author} {\bibfnamefont {A.~J.~H.}\ \bibnamefont {andK. Crozier}}, \bibinfo {author} {\bibfnamefont {J.}~\bibnamefont {Aizpurua}},\ and\ \bibinfo {author} {\bibfnamefont {R.}~\bibnamefont {Hillenbran}},\ }\bibfield  {title} {\bibinfo {title} {Controlling the near-field oscillations of loaded plasmonic nanoantennas},\ }\href@noop {} {\bibfield  {journal} {\bibinfo  {journal} {Nature Photon.}\ }\textbf {\bibinfo {volume} {3}},\ \bibinfo {pages} {287} (\bibinfo {year} {2009})}\BibitemShut {NoStop}%
\bibitem [{\citenamefont {Schuller}\ \emph {et~al.}(2010)\citenamefont {Schuller}, \citenamefont {Barnard}, \citenamefont {Cai}, \citenamefont {Jun}, \citenamefont {White},\ and\ \citenamefont {Brongersma}}]{lsp6}%
  \BibitemOpen
  \bibfield  {author} {\bibinfo {author} {\bibfnamefont {J.~A.}\ \bibnamefont {Schuller}}, \bibinfo {author} {\bibfnamefont {E.~S.}\ \bibnamefont {Barnard}}, \bibinfo {author} {\bibfnamefont {W.}~\bibnamefont {Cai}}, \bibinfo {author} {\bibfnamefont {Y.~C.}\ \bibnamefont {Jun}}, \bibinfo {author} {\bibfnamefont {J.~S.}\ \bibnamefont {White}},\ and\ \bibinfo {author} {\bibfnamefont {M.~L.}\ \bibnamefont {Brongersma}},\ }\bibfield  {title} {\bibinfo {title} {Plasmonics for extreme light concentration and manipulation},\ }\href@noop {} {\bibfield  {journal} {\bibinfo  {journal} {Nature Mater.}\ }\textbf {\bibinfo {volume} {9}},\ \bibinfo {pages} {193} (\bibinfo {year} {2010})}\BibitemShut {NoStop}%
\bibitem [{\citenamefont {Atwater}\ and\ \citenamefont {Polman}(2010)}]{lsp7}%
  \BibitemOpen
  \bibfield  {author} {\bibinfo {author} {\bibfnamefont {H.~A.}\ \bibnamefont {Atwater}}\ and\ \bibinfo {author} {\bibfnamefont {A.}~\bibnamefont {Polman}},\ }\bibfield  {title} {\bibinfo {title} {Plasmonics for improved photovoltaic devices},\ }\href@noop {} {\bibfield  {journal} {\bibinfo  {journal} {Nature Mater.}\ }\textbf {\bibinfo {volume} {9}},\ \bibinfo {pages} {205} (\bibinfo {year} {2010})}\BibitemShut {NoStop}%
\bibitem [{\citenamefont {Pors}\ \emph {et~al.}(2012)\citenamefont {Pors}, \citenamefont {Moreno}, \citenamefont {Martin-Moreno}, \citenamefont {Pendry},\ and\ \citenamefont {Garcia-Vidal}}]{lsp8}%
  \BibitemOpen
  \bibfield  {author} {\bibinfo {author} {\bibfnamefont {A.}~\bibnamefont {Pors}}, \bibinfo {author} {\bibfnamefont {E.}~\bibnamefont {Moreno}}, \bibinfo {author} {\bibfnamefont {L.}~\bibnamefont {Martin-Moreno}}, \bibinfo {author} {\bibfnamefont {J.~B.}\ \bibnamefont {Pendry}},\ and\ \bibinfo {author} {\bibfnamefont {F.~J.}\ \bibnamefont {Garcia-Vidal}},\ }\bibfield  {title} {\bibinfo {title} {Localized spoof plasmons arise while texturing closed surfaces},\ }\href {https://doi.org/10.1103/PhysRevLett.108.223905} {\bibfield  {journal} {\bibinfo  {journal} {Phys. Rev. Lett.}\ }\textbf {\bibinfo {volume} {108}},\ \bibinfo {pages} {223905} (\bibinfo {year} {2012})}\BibitemShut {NoStop}%
\bibitem [{\citenamefont {Kern}(2009)}]{amc0}%
  \BibitemOpen
  \bibfield  {author} {\bibinfo {author} {\bibfnamefont {D.~J.}\ \bibnamefont {Kern}},\ }\href@noop {} {\emph {\bibinfo {title} {Advancements in artificial magnetic conductor design for improved performance and antenna applications}}}\ (\bibinfo  {publisher} {The Pennsylvania State University},\ \bibinfo {year} {2009})\BibitemShut {NoStop}%
\bibitem [{\citenamefont {Feresidis}\ \emph {et~al.}(2005)\citenamefont {Feresidis}, \citenamefont {Goussetis}, \citenamefont {Wang},\ and\ \citenamefont {Vardaxoglou}}]{amc1}%
  \BibitemOpen
  \bibfield  {author} {\bibinfo {author} {\bibfnamefont {A.~P.}\ \bibnamefont {Feresidis}}, \bibinfo {author} {\bibfnamefont {G.}~\bibnamefont {Goussetis}}, \bibinfo {author} {\bibfnamefont {S.}~\bibnamefont {Wang}},\ and\ \bibinfo {author} {\bibfnamefont {J.~Y.~C.}\ \bibnamefont {Vardaxoglou}},\ }\bibfield  {title} {\bibinfo {title} {Artificial magnetic conductor surfaces and their application to low-profile high-gain planar antennas},\ }\href@noop {} {\bibfield  {journal} {\bibinfo  {journal} {IEEE Transactions on Antennas and Propagation}\ }\textbf {\bibinfo {volume} {53}},\ \bibinfo {pages} {209} (\bibinfo {year} {2005})}\BibitemShut {NoStop}%
\bibitem [{\citenamefont {Ma}\ \emph {et~al.}(1998)\citenamefont {Ma}, \citenamefont {Hirose}, \citenamefont {Yang}, \citenamefont {Qian},\ and\ \citenamefont {Itoh}}]{amc2}%
  \BibitemOpen
  \bibfield  {author} {\bibinfo {author} {\bibfnamefont {K.~P.}\ \bibnamefont {Ma}}, \bibinfo {author} {\bibfnamefont {K.}~\bibnamefont {Hirose}}, \bibinfo {author} {\bibfnamefont {F.~R.}\ \bibnamefont {Yang}}, \bibinfo {author} {\bibfnamefont {Y.}~\bibnamefont {Qian}},\ and\ \bibinfo {author} {\bibfnamefont {T.}~\bibnamefont {Itoh}},\ }\bibfield  {title} {\bibinfo {title} {Realization of magnetic conducting surface using novel photonic bandgap structure},\ }\href@noop {} {\bibfield  {journal} {\bibinfo  {journal} {Electron. Lett.}\ }\textbf {\bibinfo {volume} {34}},\ \bibinfo {pages} {2041} (\bibinfo {year} {1998})}\BibitemShut {NoStop}%
\bibitem [{\citenamefont {Sievenpiper}\ \emph {et~al.}(1999)\citenamefont {Sievenpiper}, \citenamefont {Zhang}, \citenamefont {Broas}, \citenamefont {Alexopolous},\ and\ \citenamefont {Yablonovitch}}]{amc3}%
  \BibitemOpen
  \bibfield  {author} {\bibinfo {author} {\bibfnamefont {D.}~\bibnamefont {Sievenpiper}}, \bibinfo {author} {\bibfnamefont {L.}~\bibnamefont {Zhang}}, \bibinfo {author} {\bibfnamefont {R.~F.~J.}\ \bibnamefont {Broas}}, \bibinfo {author} {\bibfnamefont {N.~G.}\ \bibnamefont {Alexopolous}},\ and\ \bibinfo {author} {\bibfnamefont {E.}~\bibnamefont {Yablonovitch}},\ }\bibfield  {title} {\bibinfo {title} {High-impedance electromagnetic surfaces with a forbidden frequency band},\ }\href@noop {} {\bibfield  {journal} {\bibinfo  {journal} {IEEE Trans. Microw. Theory Tech.}\ }\textbf {\bibinfo {volume} {47}},\ \bibinfo {pages} {2059} (\bibinfo {year} {1999})}\BibitemShut {NoStop}%
\bibitem [{\citenamefont {Erentok}\ \emph {et~al.}(2005)\citenamefont {Erentok}, \citenamefont {Luljak},\ and\ \citenamefont {Ziolkowski}}]{amc4}%
  \BibitemOpen
  \bibfield  {author} {\bibinfo {author} {\bibfnamefont {A.}~\bibnamefont {Erentok}}, \bibinfo {author} {\bibfnamefont {P.}~\bibnamefont {Luljak}},\ and\ \bibinfo {author} {\bibfnamefont {R.}~\bibnamefont {Ziolkowski}},\ }\bibfield  {title} {\bibinfo {title} {Characterization of a volumetric metamaterial realization of an artificial magnetic conductor for antenna applications},\ }\href@noop {} {\bibfield  {journal} {\bibinfo  {journal} {IEEE Transactions on Antennas and Propagation}\ }\textbf {\bibinfo {volume} {53}},\ \bibinfo {pages} {160} (\bibinfo {year} {2005})}\BibitemShut {NoStop}%
\bibitem [{\citenamefont {Rayno}\ \emph {et~al.}(2014)\citenamefont {Rayno}, \citenamefont {Iskander},\ and\ \citenamefont {Celik}}]{amc5}%
  \BibitemOpen
  \bibfield  {author} {\bibinfo {author} {\bibfnamefont {J.}~\bibnamefont {Rayno}}, \bibinfo {author} {\bibfnamefont {M.~F.}\ \bibnamefont {Iskander}},\ and\ \bibinfo {author} {\bibfnamefont {N.}~\bibnamefont {Celik}},\ }\bibfield  {title} {\bibinfo {title} {Synthesis of broadband true-3d metamaterial artificial magnetic conductor ground planes using genetic programming},\ }\href@noop {} {\bibfield  {journal} {\bibinfo  {journal} {IEEE Transactions on Antennas and Propagation}\ }\textbf {\bibinfo {volume} {62}},\ \bibinfo {pages} {5732} (\bibinfo {year} {2014})}\BibitemShut {NoStop}%
\end{thebibliography}
\end{document}